\documentclass[aps, notitlepage, preprintnumbers, reprint, superscriptaddress]{revtex4-1}
\pdfoutput=1
\usepackage{graphicx,amsmath,hyperref}
\usepackage[utf8]{inputenc}
\def\o{^{\text{obs}}}

\newcommand{\fNL}{f_{\rm NL}}

\newcommand{\tpc}{(2\pi)^3}
\newcommand{\nn}{\nonumber}
\newcommand{\abs}[1]{\left\lvert#1\right\rvert}

\newcommand{\xv}{\mathbf{x}}
\newcommand{\kv}{\mathbf{k}}
\newcommand{\qv}{\mathbf{q}}
\newcommand{\rv}{\mathbf{r}}
\newcommand{\pv}{\mathbf{p}}

\newcommand{\PPhiS}{P_{\Phi, S}(k, \xv)}

\newcommand{\DeltaTT}{\frac{\Delta T}{T}}
\newcommand{\DeltaTTfNL}{\left.\DeltaTT \right|_{\fNL}}
\newcommand{\DeltaTTgaus}{\left. \DeltaTT \right|_{\rm gaus}}

\newcommand{\gNL}{g_{\rm NL}}

\newcommand{\g}{g}
\newcommand{\gLM}{\g_{LM}}

\newcommand{\lv}{V_L}
\newcommand{\rcmb}{r_{\rm cmb}}

\newcommand{\Cl}{C_{\ell}}
\newcommand{\jl}{j_{\ell}}
\newcommand{\Nextra}{N_{\rm extra}}
\newcommand{\curv}{\sigma}
\newcommand{\fNLc}{{\fNL}_{\curv}}

\newcommand{\smono}{{\sigma^{\rm mono}}}
\newcommand{\sfNLmono}{{\sigma_{\fNL}^{\rm mono}}}
\newcommand{\sGmono}{{\sigma_G^{\rm mono}}}

\newcommand{\diracdelta}{{\delta_D}}

\def\be{\begin{equation}}
\def\ee{\end{equation}}
\def\ba{\begin{eqnarray}}
\def\ea{\end{eqnarray}}

\begin{document}
\preprint{IGC-15/8-1}
\title{Large-scale anomalies in the cosmic microwave background\\ as signatures of non-Gaussianity}
\date{\today}

\author{Saroj Adhikari}
\email{adh.saroj@psu.edu}
\affiliation{Institute for Gravitation and the Cosmos, The Pennsylvania State University, University Park,\\ Pennsylvania 16802, USA}
\author{Sarah Shandera}
\email{shandera@gravity.psu.edu}
\affiliation{Institute for Gravitation and the Cosmos, The Pennsylvania State University, University Park,\\ Pennsylvania 16802, USA}
\affiliation{Perimeter Institute for Theoretical Physics, Waterloo, Ontario N2L 2Y5, Canada}
\author{Adrienne L. Erickcek}
\email{erickcek@physics.unc.edu}
\affiliation{Department of Physics and Astronomy, University of North Carolina at Chapel Hill, Phillips Hall CB 3255, Chapel Hill, North Carolina 27599, USA}

\begin{abstract} 
We derive a general expression for the probability of observing deviations from statistical isotropy in the cosmic microwave background (CMB) if the primordial fluctuations are non-Gaussian and extend to superhorizon scales. The primary motivation is to properly characterize the monopole and dipole modulations of the primordial power spectrum that are generated by the coupling between superhorizon and subhorizon perturbations. Unlike previous proposals for generating the hemispherical power asymmetry, we do not assume that the power asymmetry results from a single large superhorizon mode. Instead, we extrapolate the observed power spectrum to superhorizon scales and compute the power asymmetry that would result from a specific realization of non-Gaussian perturbations on scales larger than the observable universe. Our study encompasses many of the scenarios that have been put forward as possible explanations for the CMB hemispherical power asymmetry. We confirm our analytic predictions for the probability of a given power asymmetry by comparing them to numerical realizations of CMB maps. We find that non-local models of non-Gaussianity and scale-dependent local non-Gaussianity produce scale-dependent modulations of the power spectrum, thereby potentially producing both a monopolar and a dipolar power modulation on large scales. We then provide simple examples of finding the posterior distributions for the parameters of the bispectrum from the observed monopole and dipole modulations. 
\end{abstract}

\maketitle

\section{Introduction}
It is very tempting to try to use large-scale features of the primordial fluctuations, such as the hemispherical power asymmetry, as a clue toward primordial physics. Such signals are both intriguing (maybe they say something about the beginning of inflation) and statistically unfortunate (i.e., not all that unlikely to be a feature of a particular realization of Gaussian, isotropic fluctuations). Many studies of the observed power asymmetry and its statistical significance have been reported using the WMAP data \cite{2004ApJ...605...14E, 2009ApJ...699..985H, Bennett:2010jb} and the {\it Planck} data \cite{Ade:2013nlj, Flender:2013jja, Akrami:2014eta, Quartin:2014yaa, Ade:2015hxq, Aiola:2015rqa}. A variety of possible explanations for this asymmetry have been discussed (see e.g. \cite{Dai:2013kfa}): several of the most intriguing ideas use superhorizon fluctuations to generate the asymmetry, either by using non-Gaussianity to couple them to observable perturbations \cite{Erickcek:2008sm, Erickcek:2009at, Schmidt:2012ky, Lyth:2013vha, Kanno:2013ohv, Wang:2013lda, D'Amico:2013iaa, Kohri:2013kqa, 2013PhRvL.111k1302L, McDonald:2013aca, McDonald:2013qca, Mazumdar:2013yta, Namjoo:2013fka, Abolhasani:2013vaa, Assadullahi:2014pya, Firouzjahi:2014mwa, Jazayeri:2014nya, McDonald:2014lea, Namjoo:2014nra, Lyth:2014mga, Agullo:2015aba, Byrnes:2015asa, Kenton:2015jga, Kobayashi:2015qma} or by postulating some different primordial physics that precedes the usual slow-roll inflation \cite{Donoghue:2007ze, Liu:2013kea,Liu:2013iha, Ashoorioon:2015pia}. 
Alternatively, one can postulate scenarios that are fundamentally anisotropic on the largest scales \cite{2011PhRvD..84b3014M, Cai:2013gma, Chang:2013lxa, Chang:2013mya, Kothari:2015tqa}.

It is quite general that if the primordial fluctuations are non-Gaussian, the likelihood of observing statistical anisotropies in our cosmic microwave background (CMB) changes. Although one might expect that isotropy and Gaussianity are independent criteria for the statistics of the primordial fluctuations, this distinction is not actually clear when we only have access to a finite volume of the universe \cite{Ferreira:1997wd, Lewis:2011au, Pearson:2012ba}. If the primordial fluctuations are non-Gaussian, the observed large-scale ``discrepancies" from the simplest isotropic, power law, Gaussian fluctuations need not be a signal of a special scale, time, or feature during the primordial (inflationary) era. Instead, they may simply be a consequence of cosmic variance in a universe larger than the volume we currently observe, filled with non-Gaussian fluctuations.  

In this paper, we present a single framework to calculate the distribution of expected deviations from isotropy in our observed sky from any model with non-Gaussian primordial fluctuations. This framework incorporates most successful proposals for generating the power asymmetry, even some that were not originally formulated as non-Gaussian models. The reason is that if the assumption of statistical isotropy is maintained, any explanation of the temperature power asymmetry can be modeled by assuming a fluctuation in a long wavelength modulating field that couples to some cosmological parameter relevant for determining the CMB power spectrum (the fundamental constants, the scalar spectral index and the inflaton decay rate, for example \cite{2011PhRvD..84b3014M, Dai:2013kfa, Cai:2013gma, McDonald:2013qca, McDonald:2014lea, Zarei:2014raa}). Since the asymmetry is well modeled by a (small) range of shorter scales all coupling to a longer scale this is, by definition, a sort of non-Gaussianity. 

Using our analytic framework together with numerical realizations, we show that isotropy violations in our observable universe are more likely in models where subhorizon modes are coupled to longer wavelength modes. In some models they are so much more likely that the extent to which our observed sky is isotropic can be interpreted as a constraint on non-Gaussianity. 

In addition, we find that scale dependence of such modulations is a generic outcome of non-Gaussianity beyond the standard local ansatz. This result is important because a major hurdle in model building for the power anomalies has been their scale dependence, as the observations suggest a very sharp decrease in the power asymmetry amplitude at smaller scales \cite{Flender:2013jja, Hirata:2009ar, Adhikari:2014mua}. Scale dependence of the type observed occurs whenever modes closer to the present Hubble scale couple more strongly to superhorizon modes than very short wavelength modes do. This can be the case even for scale-invariant bispectra. For example, since equilateral-type non-Gaussianity peaks when the momenta configurations are nearly equal, the modulation from an equilateral-type bispectrum is biggest at large scales and quickly dies off at smaller CMB scales.

Further, we will see that any non-Gaussianity used to explain the scale-dependent power asymmetry will also produce a scale-dependent modulation of the power-spectrum monopole. This means that the observed low power on large scales and the dipole power asymmetry can both provide evidence in favor of non-Gaussian fluctuations. To quantify the significance of these signals, we perform a parameter estimation of the non-Gaussian amplitude and shape using both monopole and dipole modulations, and estimate Bayesian evidences against the Gaussian model.

Finally, we will show that scale-dependent non-Gaussianity eliminates the need for enhanced inhomogeneity on superhorizon scales to generate the observed asymmetry.  The most common method for introducing a dipolar power modulation is to postulate the existence of a large-amplitude superhorizon fluctuation in a spectator field during inflation that then alters the power spectrum on smaller scales via local-type non-Gaussianity. (A multiple-field model of inflation, or one that otherwise breaks the usual consistency relation, is required because superhorizon perturbations in the standard inflaton field cannot generate an asymmetry \cite{Mirbabayi:2014hda}.)  Dubbed the Erickcek-Kamionkowski-Carroll (EKC) mechanism \cite{Erickcek:2008sm}, this approach has been expanded upon and refined several times since its inception \cite{Erickcek:2009at, Abolhasani:2013vaa, D'Amico:2013iaa, Kanno:2013ohv, Kohri:2013kqa, 2013PhRvL.111k1302L, Lyth:2013vha, Mazumdar:2013yta, McDonald:2013aca, Namjoo:2013fka, Wang:2013lda, Assadullahi:2014pya, Firouzjahi:2014mwa, Jazayeri:2014nya, Lyth:2014mga, McDonald:2014lea, Namjoo:2014nra, Agullo:2015aba, Byrnes:2015asa, Kenton:2015jga, Kobayashi:2015qma}.  In these analyses, WMAP and Planck bounds on local-type non-Gaussianity forced the amplitude of the superhorizon perturbation to be much larger than predicted by an extrapolation of the observed primordial power spectrum to larger scales.  Although possible origins for this large-amplitude fluctuation have been proposed, such as a supercurvature perturbation in an open universe \cite{ 2013PhRvL.111k1302L}, a bounce prior to inflation \cite{Agullo:2015aba}, a non-vacuum initial state \cite{Firouzjahi:2014mwa}, topological defects \cite{Kohri:2013kqa, Jazayeri:2014nya}, or deviations from slow-roll inflation \cite{Mazumdar:2013yta,McDonald:2013aca}, it is largely taken as an \textit{ad hoc} addition to the inflationary landscape.  The scale dependence of the asymmetry often requires an additional elaboration to the theory, either in the form of scale-dependent non-Gaussianity \cite{Erickcek:2009at, Lyth:2013vha, Kohri:2013kqa, Firouzjahi:2014mwa, Lyth:2014mga, Byrnes:2015asa, Agullo:2015aba} or isocurvature fluctuations \cite{Erickcek:2009at, McDonald:2013aca, Assadullahi:2014pya}.  

Dropping the idea of a special large superhorizon fluctuation and instead starting with scale-dependent non-Gaussianity changes this picture. Importantly, the constraints on the amplitude of the non-Gaussianity on large scales is rather weak; the Planck bound of $\fNL^{\rm local}=2.5\pm5.7$ at 68\% CL \cite{Ade:2015ava} assumes a scale-invariant bispectrum. To generate the observed power asymmetry, we only need non-Gaussianity on the scales that are asymmetric. For example, if we consider CMB multipoles up to $\ell \simeq 100$, the WMAP 5 year data give $\fNL=-100\pm100$ \cite{Smith:2009jr}. (This constraint has not appreciably changed since WMAP5: see also Figure 11 of \cite{Ade:2015ava} for the most recent plot of Planck's $\ell_{\rm max}$-dependent constraints.) Many arbitrary choices for $\ell_{\rm max}$ in the range $[40, 600]$ (and different binning schemes to study the scale dependence) can be found in the literature for computing the large-scale power asymmetry amplitude and its significance \cite{2004ApJ...605...14E, Flender:2013jja, Ade:2013nlj, Ade:2015hxq}. We will use $\ell_{\rm max}=100$ as our fiducial value to compute the asymmetry for our numerical tests and statistical analysis later. This also allows us to use the WMAP5 large-scale bispectrum constraints computed with $\ell_{\rm max}=100$. 

The weaker constraint on $\fNL$ implies that it is possible to generate the observed asymmetry without enhancing the amplitude of superhorizon fluctuations, as was noted by \cite{Schmidt:2012ky}, which considered the power-spectrum modulations generated by anisotropic bispectra.   We note, however, that the dipolar bispectrum considered there does not generate a power asymmetry because the $\vec{k}\rightarrow-\vec{k}$ symmetry of the power spectrum forbids any modulation to the power spectrum from bispectra that depend on odd powers of the angle between the long and short modes; this point was clarified in \cite{Schmidt:2015xka}. In this work, we will focus on isotropic, scale-dependent non-Gaussianity (although our framework can easily be extended to include fundamentally anisotropic models), and we will show that this is sufficient to generate the observed asymmetry without enhancing the amplitude of superhorizon fluctuations.  

The plan of the paper is as follows. In the next section, we use the usual local ansatz to demonstrate the validity of our analytic calculations of the statistical anisotropies expected in non-Gaussian scenarios and to illustrate several of the key conceptual points relating non-Gaussianity and anisotropies.  We also show how our framework encompasses the EKC mechanism and demonstrate that exotic superhorizon perturbations are not required to generate the observed power asymmetry. In Section \ref{sec:numandstat}, we test our analytic calculations for monopole and dipole power modulations using numerical realizations of CMB maps. We also introduce and discuss our parameter estimation and model comparison methods.  In Section \ref{sec:goodModel}, we investigate non-Gaussianity beyond the local ansatz and in particular, consider a representative model that generates features that are fully consistent with constraints on the isotropic power spectrum and bispectrum. We discuss and summarize important aspects of our work and conclude in Section \ref{sec:conc}. The appendixes contain technical details.

\section{Illustrating the connection between non-Gaussianity and isotropy}
\label{sec:localModel}
We assume that at some early time (after reheating but prior to the release of the cosmic microwave background radiation) a large volume of the Universe ($V_L$) contains adiabatic fluctuations described by isotropic but non-Gaussian statistics. To compare predictions of this model with observations, we are interested in the statistics of the fluctuations in smaller volumes, $V_S\ll V_L$, that correspond in size to our presently observable Hubble volume. We will first consider the usual local model with constant $\fNL$ for simplicity; we will present more general results in a later section.

\subsection{The local model}
Suppose the Bardeen potential $\Phi$ is a non-Gaussian field described by the local model: 
\begin{eqnarray}
\Phi(\xv) = \phi(\xv) + \fNL\left(\phi(\xv)^2- \langle \phi(\xv)^2\rangle \right),
\label{eq:localModel}
\end{eqnarray}
where $\phi(\xv)$ is a Gaussian random field. When the large volume is only weakly non-Gaussian, the power spectrum observed in our sky, $\PPhiS$, will be related to the mean power spectrum in the large volume, $P_\phi(k)$, by
\begin{eqnarray}
 \PPhiS &=& P_\phi(k) \left[1 + 4 \fNL \int \frac{d^3 \kv_{\ell}}{\tpc} \phi(\kv_{\ell}) e^{i\kv_{\ell} \cdot \xv} \right], \nn \\
 \label{eq:modulatedpowerLocal}
\end{eqnarray}
where the radial integration for $k_{\ell}$ is confined to \mbox{$|k_{\ell}|<\pi/\rcmb$} when the CMB spectrum is the quantity of interest. Our conventions for the power spectrum are stated in Appendix \ref{app:conventions}, and Appendix \ref{app:derivation} provides the derivation of this equation. Any particular model for generating the fluctuations in volume $V_L$ should provide a well-motivated lower bound on the $k_{\ell}$ integral (e.g., from the duration of inflation).  However, not all shifts to local statistics are sensitive to the full range of the integral; for local-type non-Gaussianity, as we will show later, only the monopole receives contributions from all super-Hubble modes.  

The power spectrum $P_{\phi}$ and amplitude of non-Gaussianity, $\fNL$, appearing on the right-hand side of Eq.(\ref{eq:modulatedpowerLocal}) are those defined in the large volume. However, looking ahead to the result for the dipole modulation from the local ansatz, Eq.(\ref{eq:dipole}), $P_{\phi}$ and $\fNL$ will be shifted to the observed values. In particular, the local non-Gaussianity that generates the observed power asymmetry is only the portion that violates Maldacena's consistency relation \cite{Maldacena:2002vr} and is zero in single-clock inflation \cite{Mirbabayi:2014hda}. Since the observed values are ultimately the relevant quantities in the analysis, we will not increase the complexity of the notation to distinguish the large and small volume parameters, except in the appendixes.

The field $\phi(\kv_{\ell})$ appearing in the integral is no longer stochastic but consists of the particular realization of the field that makes up the background of a particular Hubble volume.  The power spectrum in Eq.(\ref{eq:modulatedpowerLocal}) can depend on position $\xv$ within $V_S$ because an individual realization (local value) of the fluctuations $\phi(\kv_{\ell})$ can be non-zero. If we consider the average statistics in the large volume (equivalent to averaging over all regions of size $V_S$), then the term proportional to $\fNL$ in Eq.(\ref{eq:modulatedpowerLocal}) above averages to zero since $\langle \phi(\kv_{\ell}) \rangle _{\lv}=0$. In that case we recover the isotropic power spectrum of $V_L$. Finally, keep in mind that Eq.(\ref{eq:modulatedpowerLocal}) is still more general than our actual CMB sky: it provides the statistics from which to draw realizations of our observed modes. Any single sky realization will still be subject to the usual cosmic variance that affects the values of small $\ell$ modes and that can generate a power asymmetry even for volumes where the term proportional to $\fNL$ is zero. 

It is also interesting to consider a two-field extension of Eq.(\ref{eq:localModel}):
\begin{equation}
 \Phi(\xv) = \varphi(\xv) + \sigma(\xv) + \fNLc \left( \curv(\xv)^2 -\langle \curv(\xv)^2 \rangle \right),
 \label{eq:twofieldlocalmodel}
\end{equation}
where both $\varphi(\xv)$ and $\sigma(\xv)$ are Gaussian random fields and are uncorrelated. In this case, the power spectrum observed in our sky, $P_{\Phi, s}(k, \xv)$, is
\begin{equation}
 \PPhiS = P_\Phi(k) \left[1 + 4 \xi \fNLc \int \frac{d^3 \kv_{\ell}}{\tpc} \curv(\kv_{\ell}) e^{i\kv_{\ell} \cdot \xv} \right],
 \label{eq:infcurvpower} 
\end{equation}
where $P_\Phi(k)=P_\varphi(k)+P_\curv(k)$ is the mean power spectrum in the large volume and $\xi = P_{\Phi, \sigma}/P_{\Phi}$ is the fraction of power in the $\curv$ field. We will only consider cases with weak non-Gaussianity ($\fNLc^2 \mathcal{P}_\curv \ll 1$ and $P_{\Phi, \curv} \approx P_{\curv}$). The amplitude of the local-type bispectrum for this weakly non-Gaussian two-field model is given by $\fNL=\xi^2 \fNLc$. Therefore, the inhomogeneous power spectrum in terms of the observed $\fNL$ and the fraction of power $\xi$ is
\begin{equation}
\PPhiS = P_\Phi(k) \left[1 + 4 \frac{\fNL}{\xi} \int \frac{d^3 \kv_{\ell}}{\tpc} \curv(\kv_{\ell}) e^{i\kv_{\ell} \cdot \xv} \right].
\label{eq:modulatedpowerTwoField} 
\end{equation}
A scale-dependent power fraction $\xi(k)$ is a natural way to generate a scale-dependent power asymmetry, as can be seen from the inhomogeneous term in Eq.(\ref{eq:infcurvpower}); if $\xi(k)$ decreases for large values of $k$, the modulation of the power spectrum will decrease as well.  Such mixed-perturbation models also have other potentially observable consequences: $\xi$ affects the tensor-to-scalar ratio and contributes to large-scale stochasticity in the power spectra of galaxies \cite{Tseliakhovich2010}.

\subsection{Effect on the CMB sky}
\label{sec:cmb}

The imprint of the inhomogeneous power spectrum given by Eq.(\ref{eq:modulatedpowerLocal}) on the CMB can be described in terms of a multipole expansion:
\begin{eqnarray}
 P_\Phi(k, \hat{n}) &=& P_\phi(k)\left[1 + \fNL \sum_{LM} \gLM Y_{LM}(\hat{n}) \right],
 \label{eq:PPhimultipole}
\end{eqnarray}
where $Y_{LM}$ is a spherical harmonic, and $\hat{n}$ is the direction of observation on the last scattering surface.  To find the expansion coefficients $g_{LM}$ we make use of the plane wave expansion
\begin{eqnarray}
 e^{i\kv_{\ell}\cdot \xv} &=& 4 \pi \sum_{LM} i^L j_L(k_{\ell} x) Y_{LM}^*(\hat{k_{\ell}}) Y_{LM}(\hat{n}),
\end{eqnarray}
where $j_L$ is a spherical Bessel function of the first kind, and $\xv = x\hat{n}$ specifies the position of the observed fluctuation: for the CMB, $x=\rcmb$ is the comoving distance to the last scattering surface.  Eq.(\ref{eq:modulatedpowerLocal}) then implies that 
\begin{equation}
 \gLM = 16\pi i^L \int_{|\kv_{\ell}|<\pi/x}  \frac{d^3\kv_{\ell}}{\tpc} j_L(k_{\ell} x) \phi(\kv_{\ell}) Y_{LM}^*(\hat{k_{\ell}}).
 \label{eq:gLM}
\end{equation}

The quantity $\gLM$ has a fixed value in any single volume $V_S$, but when averaged over all small volumes in $V_L$, $\langle \gLM \rangle_{V_L}=0$. The expected covariance, on the other hand, is non-zero:
\begin{widetext}
\begin{eqnarray}
\label{eq:variance_gLM}
 \langle \gLM \g_{L'M'}^*\rangle_{\lv} &=& 256 \pi^2 (-1)^{L'} i^{L+L'} \int_{|\kv_{\ell}|<\pi/x}\frac{d^3 \kv_{\ell}}{\tpc} j_L(k_{\ell} x) j_{L'}(k_{\ell} x) P_\phi(k_{\ell})  Y_{LM}^*(\hat{k_{\ell}}) Y_{L'M'}(\hat{k_{\ell}}); \\
 &=& \frac{32}{\pi} \delta_{LL'} \delta_{MM'} \int_0^{\pi/x} dk_{\ell}\; k_{\ell}^2 j_L^2(k_{\ell} x) P_\phi(k_{\ell}); \nonumber \\
 &=& 64 \pi \delta_{LL'} \delta_{MM'} \int_0^{\pi/x} \frac{dk_{\ell}}{k_{\ell}} j_L^2(k_{\ell} x) \mathcal{P}_\phi(k_{\ell}),
\end{eqnarray}
\end{widetext}
where in the last line, we have defined the dimensionless power spectrum as $\mathcal{P}_\phi(k) = k^3P_\phi(k)/(2\pi^2)$. We have again used the subscript $\lv$ to indicate the ensemble average is over the values of $g_{LM}$ in the full volume $V_L$. Note that both the individual values of $g_{LM}$ and their variance depend on the size of the small volume through the upper limit of integration in Eqs. (\ref{eq:gLM}) and (\ref{eq:variance_gLM}). While the mean statistics in the large volume cannot depend on the scale for the small volume, the {\it variance} of the statistics observed in sub-volumes generically does. It is now straightforward to study the monopole and dipole contributions from non-Gaussian cosmic variance to the modulated component of the power spectrum in a small volume.

In the case of the two-field extension, using Eq.(\ref{eq:modulatedpowerTwoField}) in the definition of the modulation moments Eq.(\ref{eq:PPhimultipole}) gives
\begin{eqnarray}
 \Big\langle g_{LM}g_{L'M'}^*\Big\rangle_{\lv} &=& \frac{64 \pi \delta_{LL'} \delta_{MM'}}{\xi^2} \int_0^{\pi/x} \frac{dk_{\ell}}{k_{\ell}} j_L^2(k_{\ell} x) \mathcal{P}_\curv(k_{\ell}); \nn \\
 &=& \frac{1}{\xi} \Big\langle g_{LM} g_{L'M'}^*\Big\rangle_{\lv, \xi=1}.
 \label{eq:twofieldvariance}
\end{eqnarray}
That is, for the same amplitude of non-Gaussianity observed in the $\Phi$ field, the variance of the non-Gaussian modulations increases by a factor of $1/\xi$ compared to the single source ($\xi=1$) local model.

\subsubsection{Monopole modulation ($L=0$)}
\label{subsec:cmbmono}
The power-spectrum amplitude shift, $A_0$, in the parametrization of Eq.(\ref{eq:PPhimultipole}) is: 
\begin{equation}
P_\Phi(k) = P_\phi(k) \left[1 + A_0 \right]= P_\phi(k) \left[1 + \fNL\frac{\g_{00}}{2\sqrt{\pi}} \right],
\label{eq:monopole}
\end{equation}
where $A_0$ can be either positive or negative, but has a lower bound $A_0\geq-1$. From the discussion above and Eq.(\ref{eq:variance_gLM}), it is clear that $\g_{00}$ is Gaussian distributed with zero mean and variance given by:
\begin{eqnarray}
 \langle \g_{00}^2 \rangle &=& 64\pi \int \frac{dk_{\ell}}{k_{\ell}} \left[\frac{\sin(k_{\ell} x)}{k_{\ell} x}\right]^2 \mathcal{P}_\phi(k_{\ell}).
 \label{eq:monopolevariance}
\end{eqnarray}
Therefore, the distribution of the monopole power modulation amplitude $A_0$ also follows a normal distribution, for small values of $A_0$, ($|A_0| \ll 1$), with zero mean and standard deviation:
\begin{eqnarray}
 \sfNLmono = \frac{1}{2\sqrt{\pi}} |\fNL| \langle g_{00}^2 \rangle ^{\frac{1}{2}}.
 \label{eq:sigmafNLmono}
\end{eqnarray}
The expression for $\langle \g_{00}^2 \rangle$ is sensitive to the infrared limit of the integral. That is, all super-Hubble modes can contribute. Interesting aspects of cosmic variance arising from this term, including effects on the observed non-Gaussianity in small volumes have been subjects of investigation in \cite{Linde:2005yw, Nelson:2012sb, LoVerde:2013xka, Nurmi:2013xv, LoVerde:2013dgp, Bramante:2013moa, 2015PhRvD..91h3518B}. In particular, the observed value of $\fNL$ is, in general, shifted from the mean value in the large volume. 

For a constant $\fNL$, the effect of the monopole modulation is to change the power-spectrum amplitude on all scales and therefore is not observationally distinguishable from the ``bare'' value of the power-spectrum amplitude. For scale-dependent non-Gaussianity, there is a scale-dependent power modulation, which can generically be interpreted as shifting the spectral index in the small volume away from the mean value in the large volume. In cases where the amplitude of non-Gaussianity is small (and consistent with zero) at small scales (large $\ell$), the power-spectrum amplitude from those scales can be taken as $P_\phi(k)$, and then one can look for monopole modulation at large scales for which the non-Gaussianity constraints are not as strong. The large-scale power suppression anomaly \cite{Ade:2013kta, Contaldi:2003zv} is exactly such a situation. We will return to this point in more detail in Section \ref{sec:goodModel}.

\subsubsection{Dipole modulation ($L=1$)}
The dipole modulation of the power spectrum in the parametrization of Eq.(\ref{eq:PPhimultipole}) is given by:
\begin{equation}
 P_{\Phi}(k, \hat{n}) = P_{\phi}(k) \left[ 1 + \fNL \sum_{M=-1,0,1} \g_{1M} Y_{1M}(\hat{n}) \right].
 \label{eq:dipole}
\end{equation}
Since we are interested in the dipole modulation of the observed power spectrum in the CMB sky, the above equation should be obtained from Eq.(\ref{eq:modulatedpowerLocal}) by absorbing the (unobservable) monopole shift to the observed power spectrum. Then, on the right-hand side of Eq.(\ref{eq:dipole}), $P_{\phi}(k)$ is the observed isotropic power spectrum and $\fNL$ is the observed amplitude of local non-Gausianity within our Hubble volume. See Eq. (\ref{eq:dipolePobs}) and the discussion there for details. (Appendix \ref{app:biposh} contains the corresponding expression in terms of bipolar spherical harmonics.)  
 The $\g_{1M}$ coefficients are Gaussian distributed with zero mean and a variance
 \begin{equation}
 \langle \g_{1M} \g^*_{1M} \rangle = 64 \pi \int \frac{dk_{\ell}}{k_{\ell}} \left[\frac{\sin(k_{\ell} x)}{(k_{\ell} x)^2} - \frac{\cos(k_{\ell} x)}{k_{\ell} x} \right]^2 \mathcal{P}_\phi(k_{\ell}). \nonumber
 \label{eq:dipolevariance}
\end{equation}

If we pick a direction $\vec{d}_i$ in which to measure the dipole modulation $A_i$ such that
\be
P_{\Phi}(k) = P_\phi(k) \left[1 + 2A_i \cos{\theta} \right],
\label{eq:PPhidipole}
\ee
where $\cos\theta = \hat{d_i} \cdot \hat{n}$, then the contribution to the dipole from the non-Gaussianity is $A^{\rm NG}_i = \frac{1}{4} \sqrt{\frac{3}{\pi}} \fNL \g_{10}$, which is normally distributed with mean zero and standard deviation:
\begin{eqnarray}
 \sigma_{\fNL}=\frac{1}{4}\sqrt{\frac{3}{\pi}} |\fNL| \langle g_{10}^2 \rangle^{\frac{1}{2}}
 \label{eq:sigmafNL}
\end{eqnarray}
In the two-field model Eq.(\ref{eq:twofieldlocalmodel}), using Eq.(\ref{eq:twofieldvariance}), the standard deviation gets modified:
\begin{eqnarray}
 \sigma_{\fNL}&=& \frac{1}{\sqrt{\xi}} \left[ \sigma_{\fNL}\right]_{\xi=1}\;.
\end{eqnarray}
This shows that for $\xi<1$ (e.g. a mixed inflaton-curvaton model), it is easier to generate the hemispherical power asymmetry with a small value of $\fNL$.  However, there is a minimal value of $\xi$ that can generate a power asymmetry of a given amplitude: the requirement of weak non-Gaussianity in the non-Gaussian field [$\curv(\xv)$ in Eq.(\ref{eq:modulatedpowerTwoField})] demands that $\xi \gtrsim A_i^{\rm NG}$.

The above discussion of the distribution of the dipole asymmetry $A_i$ assumes that we measure $A_i$ in a fixed direction $\vec{d_i}$. However, we have no \textit{a priori} choice of direction $\vec{d}_{i}$ in most situations. This is especially true when considering a power asymmetry that is generated by the random realization of superhorizon perturbations as opposed to a single exotic perturbation mode. 
Therefore, observations of dipole power modulations are necessarily reported using the amplitude of dipole modulation in the direction of the maximum modulation. To obtain that amplitude, we can consider any three orthonormal directions ($d_1, d_2, d_3$) on the CMB sky and measure the corresponding three dipole modulation amplitudes ($A_1, A_2, A_3$) in the three corresponding orthonormal directions for each sky. The amplitude of modulation for the CMB sky (simulated or observed) is then given by $A=(A_1^2+A_2^2+A_3^2)^{\frac{1}{2}}$. Clearly then, $A$ follows the $\chi$ distribution with three degrees of freedom (also known as the Maxwell distribution). In the Section \ref{sec:num}, we will directly test the distributions and parameters obtained in this section using numerical realizations of CMB maps.

\begin{figure}
 \includegraphics[width=0.48\textwidth]{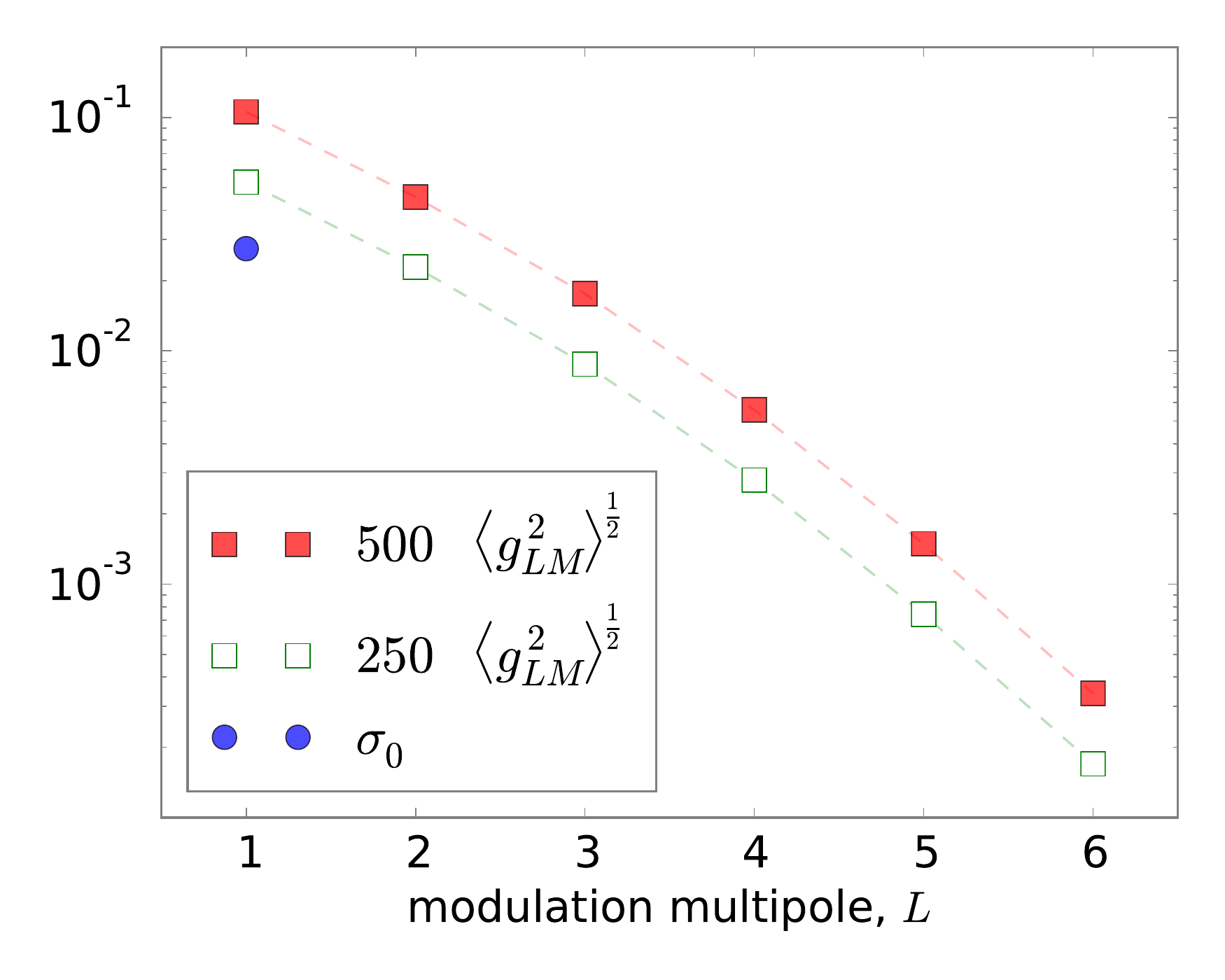}
 \caption{(color online). The expected higher order modulations for $\fNL=500, 250$ (red square, open green square). We see a sharp decrease in the expected modulation amplitude from superhorizon fluctuations at higher $L$ multipoles. The blue circle for $L=1$ is the variance measured from our Gaussian Sachs-Wolfe CMB maps. The dependence of the corresponding variances for higher multipoles $L$ in Gaussian CMB maps on the multipole $L$ is much weaker than the sharp falloff seen in the expected amplitude due to local non-Gaussianity (See Figure 2(d) of \cite{Akrami:2014eta}, for example).} 
 \label{fig:higher}
\end{figure}

\subsubsection{Higher multipole modulations}
The anisotropic modulation of the power spectrum is expected to continue to higher multipoles in the presence of non-Gaussianity. However, as shown in Figure \ref{fig:higher}, the expected value of the modulation gets smaller quickly for higher multipoles $L$. The corresponding expected variance of higher multipole modulations for Gaussian CMB maps, however, is only weakly dependent on $L$. See, for example, Figure 2(d) of \cite{Akrami:2014eta}.  There is no evidence for modulation at higher order multipoles in the Planck temperature anisotropies data (see Figure 34 of \cite{Ade:2013nlj}). In statistical analysis of the kind we discuss later in Section \ref{sec:stat}, it may, nevertheless, be useful to add higher multipole modulations (at least the quadrupole $L=2$) at large scales as it may provide increased evidence for or against non-Gaussian mode coupling. An approximate constraint on $\fNL g_{2M}$ may be obtained from the result for the Fourier space quadrupole modulation constraint in \cite{Kim:2013gka}. There are two possible scenarios: (i) the expected amplitude of the modulation is larger than that from cosmic variance in the Gaussian case; in this scenario, the lack of observation of such a modulation in the data will disfavor the non-Gaussian model that is used to explain the monopole and dipole modulations. (ii) The expected amplitude of the modulation is within the cosmic variance from the Gaussian case, in which case the data are not discriminatory for or against the non-Gaussian model.

\subsection{Connection to prior work}
Before proceeding, we pause to connect Eq.(\ref{eq:dipole}) to the EKC mechanism to better see how scale-dependent non-Gaussianity eliminates the need for enhanced perturbations on superhorizon scales.  In the EKC mechanism, a single superhorizon perturbation mode in a spectator field during inflation is responsible for generating the asymmetry; the original proposal used a curvaton field \cite{Erickcek:2008sm}, but later work extended the mechanism to any source of non-Gaussian curvature fluctuations \cite{Lyth:2013vha, Kobayashi:2015qma}. For example, consider a field $\sigma$ that generates a curvature perturbation $\zeta = {\cal N}_\sigma \delta \sigma$.  A superhorizon (SH) sinusoidal fluctuation in $\sigma$,
\be
\sigma_\mathrm{SH}(\vec{x}) = \sigma_L \cos \left(\vec{k}_L\cdot \vec{x}+\theta\right),
\label{supermode}
\ee
will generate a dipolar power asymmetry in the curvature power spectrum \cite{Kobayashi:2015qma}:
\be
P_\zeta(k, \vec{x}) = \bar{P}_\zeta(k)\left[1-\frac{12}{5}\fNL(\vec{k}_L\cdot \vec{x}){\cal N}_\sigma \sigma_L  \sin \theta\right]
\ee
to first order in $k_L x$.  In terms of the Bardeen potential, $\Phi = (3/5) \zeta$, the power asymmetry is
\be
P_\Phi(k, \vec{x}) = {P}_\phi(k)\left[1-4\fNL(\vec{k}_L\cdot \vec{x}) \Phi_L  \sin \theta\right],
\label{asym_phi}
\ee
where $\Phi_L \equiv (3/5) {\cal N}_\sigma \sigma_L$.  The Fourier transform of the superhorizon fluctuation given by Eq.~(\ref{supermode}) is
\be
\Phi_\mathrm{SH} (\vec{q}) = (2\pi)^3 \frac{\Phi_L}{2} \left[ e^{i\theta} \diracdelta(\vec{k}_L-\vec{q}) + e^{-i\theta} \diracdelta(\vec{k}_L+\vec{q})\right].
\ee
Inserting this expression into Eq.(\ref{eq:gLM}) for $g_{LM}$ implies that 
\be
g_{10} = -16 \pi \Phi_L \frac16 \sqrt{\frac3\pi} k_L x \sin \theta+ {\cal O}(k_L^2 x^2). 
\ee
With this expression for $g_{10}$, Eq.~(\ref{eq:dipole}) matches Eq.~(\ref{asym_phi}) to first order in $k_L x$.  Therefore, we see that the EKC mechanism can be described by our framework.  

For a single superhorizon mode, Eq.~(\ref{asym_phi}) implies that the non-Gaussian contribution to the dipole is
\be
A^{\rm NG} = 2 |\fNL| \Delta\Phi,
\ee
where $\Delta\Phi = \Phi_L k_L x \sin \theta$ is the variation of $\Phi$ across the surface of last scatter.  Since $k_L x < 1$, $\Phi_L > \Delta \Phi$.   The RMS amplitude of $\Phi$ values given by extrapolating the observed value of ${\cal P}_\zeta$ to larger scales is 
\be
\Phi_\mathrm{rms} \simeq \sqrt{{\cal P}_\phi},
\ee
where ${\cal P}_\phi \simeq 8 \times10^{-10}$ \cite{Ade:2015xua}.
It follows that the amplitude of the superhorizon mode $\Phi_L$ is bounded from below as
\be
\frac{\Phi_L}{\Phi_\mathrm{rms}} \gtrsim \frac{A}{2|\fNL|\sqrt{{\cal P}_\phi}}.
\label{eq:rmsrat}
\ee
If $A = 0.06$ (and is entirely due to the non-Gaussianity) and $|\fNL| < 100$, then ${\Phi_L}/{\Phi_\mathrm{rms}} > 10.6$, which implies that the superhorizon mode must be at least a 10$\sigma$ fluctuation. This is why Eq.(\ref{supermode}) was not originally considered to be part of the inflationary power spectrum, but rather a remnant of pre-inflationary inhomogeneity or a domain-wall-like feature in the curvaton field \cite{Erickcek:2008sm}.  It was then necessary to consider the imprint this enhanced superhorizon mode would leave on large-scale temperature anisotropies in the CMB through the Grishchuk-Zel'dovich (GZ) effect \cite{GZ78}.  Although the curvature perturbation generated by Eq.(\ref{supermode}) does not generate an observable dipolar anisotropy in the CMB \cite{Turner91, Erickcek:2008jp, Zibin:2008fe}, it does contribute to the quadrupole and octupole moments, and observations of these multipoles severely constrain models that employ the EKC mechanism \cite{Erickcek:2008jp, Erickcek:2008sm}.

However, if we relax our upper bound on $\fNL$ to 270 or 500, Eq.(\ref{eq:rmsrat}) indicates that a $4\sigma$ or $2\sigma$ fluctuation, respectively, could generate an asymmetry with $A=0.06$.  The odds of generating the observed asymmetry are also improved by accounting for the fact that there are three spatial dimensions, which provide three independent opportunities for a large-amplitude fluctuation.  We will see in Section \ref{sec:stat} that considering the combined contributions of several superhorizon modes and accounting for the red tilt of the primordial power spectrum further increases the probability of generating the observed asymmetry, to the point that the $p-$value for $A=0.06$ increases to greater than 0.05 for $|\fNL|\gtrsim 300$.  Thus, if the perturbations on large scales are sufficiently non-Gaussian, there is no need to invoke enhanced superhorizon perturbations to generate the observed power asymmetry.  

In the absence of an enhancement of the superhorizon power spectrum, the variance of the quadrupole moments and octupole moments in the CMB will not be altered.  Consequently, we do not expect significant constraints on such models from the GZ effect.  We note though that the specific realization of modes outside our sub-volume will still source quadrupole and octupole anisotropies in the CMB.  For realizations that generate a large power asymmetry, the GZ contribution to these anisotropies would likely be larger than expected from theoretical predictions of $C_2$ and $C_3$ and aligned with the power asymmetry.  However, this effect may be difficult to disentangle from the monopole power modulation described in Section \ref{subsec:cmbmono}, and we leave a detailed analysis of this observational signature to future work. 

Furthermore, the ratio $\Delta \Phi/\Phi_\mathrm{rms}$ for a given value of $A$ and $\fNL$ can be significantly reduced if we consider mixed Gaussian and non-Gaussian perturbations. Using the mixed perturbation scenario introduced in Eq.(\ref{eq:twofieldlocalmodel}), (with $\xi = P_{\Phi,\sigma}/P_\Phi$), 
\be
\frac{\Phi_{L,\sigma}}{\Phi_\mathrm{rms,\sigma}} > \frac{A\sqrt{\xi}}{2|\fNL| \sqrt{{\cal P}_\phi}} > \frac{A^{3/2}}{2|\fNL| \sqrt{{\cal P}_\phi}}.
\ee
In the last inequality, we employ the fact that $\xi > A$ is required for the non-Gaussianity in the $\sigma(\xv)$ field to be weak enough to make the $\mathcal{O}({\fNL}_\sigma^2 \mathcal{P}_\sigma)$ contribution to $P_{\Phi, \sigma}(k)$ negligible. In this case, $\Phi_L$ could be sourced by a 1$\sigma$ fluctuation in the $\sigma(\xv)$ field if $|\fNL|=270$. The possibility of using a mixed curvaton-inflaton model to generate a scale-dependent asymmetry using a single large superhorizon perturbation was explored in \cite{Erickcek:2009at}.

\section{Statistical Anisotropy in the CMB Power Spectrum}
\label{sec:numandstat}
\subsection{Numerical tests}
\label{sec:num}
We now present numerical tests of our analytic expressions for the dipole power modulation in the case of local non-Gaussianity. We will work in the Sachs-Wolfe (SW) regime: we only consider
\be
\DeltaTT = -\frac{\Phi}{3}.
\ee 
Therefore, for local non-Gaussianity, Eq.(\ref{eq:localModel}), the temperature fluctuation is given by:
\be
 \DeltaTTfNL = \DeltaTTgaus - 3\fNL \left[ \DeltaTTgaus^2 - \Bigg\langle \DeltaTTgaus^2 \Bigg\rangle\right]
 \label{eq:localCMB}
\ee
We generate 10000 simulated Gaussian SW CMB skies using 
\begin{eqnarray}
\label{eq:Cl}
 \Cl = \frac{4\pi}{9} \int_0^\infty \frac{dk}{k} \mathcal{P}_{\phi}(k) \jl^2(kx)
 \label{eq:SWCls}
\end{eqnarray}
for $0\leq \ell \leq300$; the primordial power spectrum is given by
\begin{eqnarray}
 \mathcal{P}_\phi(k) = A_\phi\left(\frac{k}{k_0}\right)^{n_s-1}
\end{eqnarray}
with $A_\phi=7.94\times10^{-10}$ and $n_s=0.965$ (from \textit{Planck} TT,TE,TE+lowP column in Table 3. of \cite{Ade:2015xua}), and $k_0=0.05\;{\rm Mpc^{-1}}$ as the pivot scale.

Then it is easy to generate non-Gaussian Sachs-Wolfe CMB temperature maps using Eq.(\ref{eq:localCMB}) for a constant $\fNL$. Unlike most CMB analyses, we will keep the dipole variance term $C_1$. A non-zero $C_1$ is used to model the dipolar anisotropy in density fluctuations on the scale of the observable universe (from the perspective of the large volume $\lv$). However, note that the $C_1$ we use is not what we would measure for the CMB dipole, even if we assume that the dominant contribution to the measurement of the dipole from our local motion \cite{Aghanim:2013suk} has been subtracted out. This is because, for adiabatic fluctuations, the leading-order contribution to the observed CMB dipole from superhorizon perturbations exactly cancels the Doppler dipole generated by the superhorizon perturbations \cite{Erickcek:2008jp, Zibin:2008fe}.

It is convenient to set the monopole $C_0$ to zero for the purpose of studying dipole modulations; otherwise, the cosmic variance power asymmetry (i.e. the contribution that is not due to local non-Gaussianity) will be different for the weakly non-Gaussian realization compared to the Gaussian realization from which it is generated. Therefore, in this section, we use numerical realizations with non-zero $C_0$ values only when testing the monopole modulation formula. The expression for $C_0$ is infrared divergent, so we assume an infrared cutoff $k_{\rm min}$; the same cutoff scale is used to compute the expected amount of monopole power modulation $A_0$. Numerically,
\be
 C_{0} = \frac{4\pi}{9} \int_{k_{\rm min}}^{\infty} \frac{dk}{k} \mathcal{P}_{\phi}(k) \left[ \frac{\sin{kx}}{kx} \right]^2.
 \label{eq:C0}
\ee
The $k_{\rm min}$ cutoff can be related to the number of superhorizon e-folds of inflation (if interpreted as such) as: $\Nextra=\ln\left[{(\pi/\rcmb)/k_{\rm min}}\right]$. The above integral gets most of its contribution from $k < \pi/\rcmb$, and therefore can be well approximated by \cite{LoVerde:2013xka}:
\be
 C_0 \approx \frac{4\pi}{9} A_\phi \left(\frac{\pi}{k_0\rcmb}\right)^{n_s-1}\left[ \frac{1-e^{-(n_s-1)\Nextra}}{n_s -1}\right]
 \label{eq:C0N}
\ee
for $n_s\neq1$, and $(4\pi/9) A_\phi \Nextra$ for $n_s=1$. Additional details about our numerical results can be found in Appendix \ref{app:numerics}.

\subsection{Monopole modulation ($L=0$)}
\begin{figure}
 \includegraphics[width=0.48\textwidth]{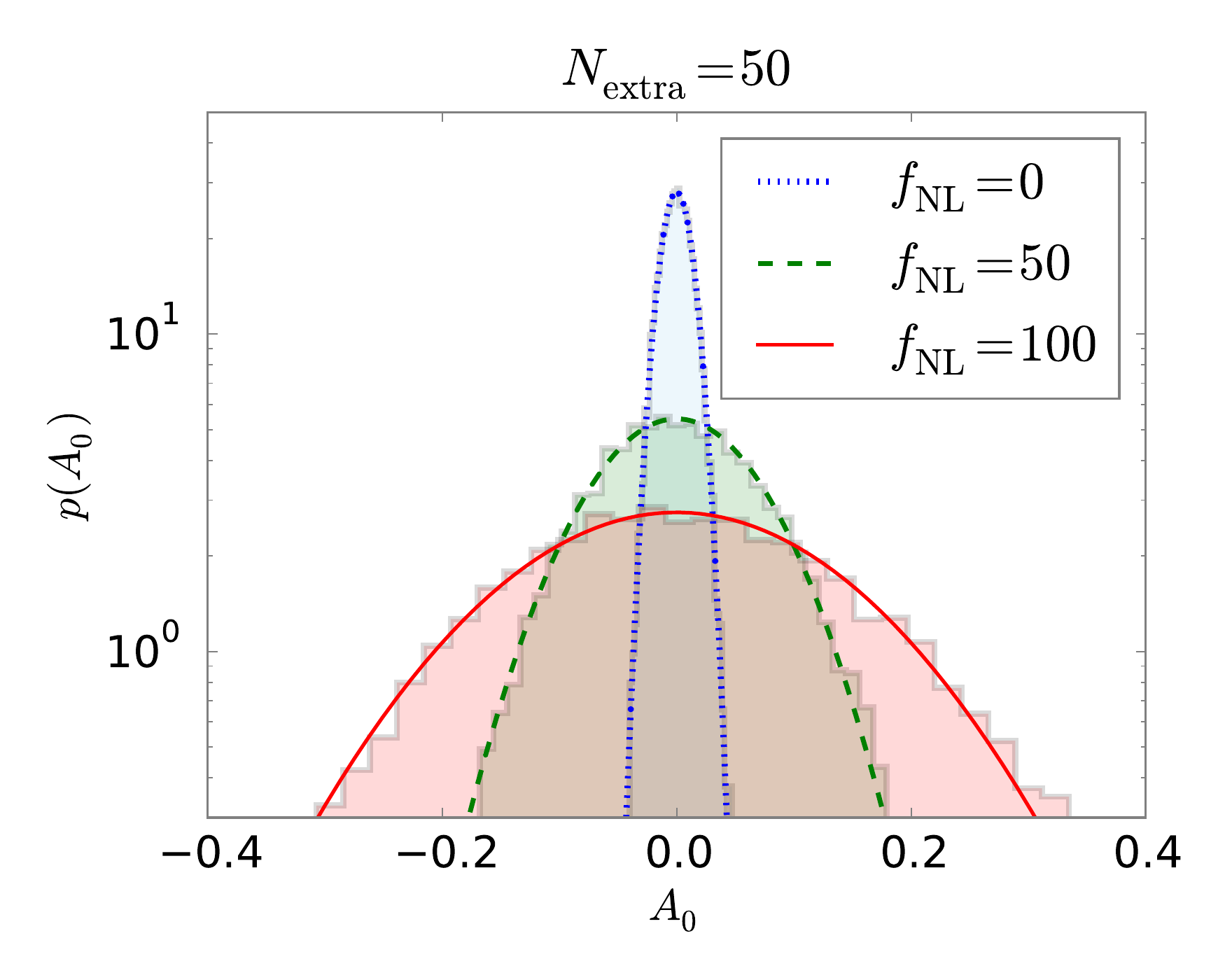}
 \caption{(color online). Test of the monopole modulation formula Eq.(\ref{eq:monopolemodulation}) for the local non-Gaussian model with $\fNL$ specified in the figure. The dotted blue line for the $\fNL=0$ (Gaussian) model is the best-fit normal distribution to the distribution obtained from numerically generated Sachs-Wolfe CMB maps, while the other two curves (dashed green, $\fNL=50$ and solid red, $\fNL=100$) are obtained using Eq.(\ref{eq:monopolemodulation}). The value of the CMB monopole $C_0$ is set using Eq.(\ref{eq:C0N}) with $\Nextra=50$. The measurement of $A_0$ is bounded below i.e. $A_0\geq-1$. The normal distribution is an excellent fit for small modulations i.e. $|A_0|\ll 1$, whereas the distribution is positively skewed for larger values of $|A_0|$.}
 \label{fig:A0fNL}
\end{figure}

The normally distributed monopole shift amplitude $A_0$ for a local non-Gaussian model is given by Eq.(\ref{eq:sigmafNLmono}), and can be written in terms of $C_0$ using Eq.(\ref{eq:C0}) as:
\be
\sfNLmono \approx 6 |\fNL| \sqrt{\frac{C_0}{\pi}}.
 \label{eq:sigmono}
\ee
The necessary infrared cutoff has already been set by the value of $\Nextra$ in Eq.(\ref{eq:C0N}) to compute $C_0$.  Although it is not possible to observe monopole modulations for a constant local $\fNL$, we can test the expected modulations assuming a value of $\Nextra$. The probability distribution of the shift for any $\fNL\neq 0$ is
\be
 p_N(A_0, \smono) = \frac{1}{\smono \sqrt{2\pi}} {\rm Exp} \left[-\frac{1}{2}\left(\frac{A_0}{\smono}\right)^2 \right], 
 \label{eq:monopolemodulation}
\ee
where the variance has contributions from the Gaussian realization and the non-Gaussian coupling to the realization of long wavelength modes: $(\smono)^2 = (\sfNLmono)^2 + (\sGmono)^2$. For our numerical tests, $\sGmono$ is the variance of $A_0$ measured in Gaussian CMB maps. The quantity we measure for $A_0$ from each realization of CMB maps is:
\be
A_0 = \frac{1}{\sum_{\ell=2}^{\ell_{\rm max}} (2\ell+1)} \sum_{\ell=2}^{\ell_{\rm max}} (2 \ell + 1) \left[\frac{\Cl -\Cl^{\rm true}}{\Cl^{\rm true}}\right],
\label{eq:A0}
\ee
where $\Cl^{\rm true}$ are the input angular power-spectrum values used to obtain the set of numerical CMB maps, and $C_\ell$ is the angular power spectrum of a particular realization of that set of CMB maps, and $\ell_{\rm max}=100$. In Figure \ref{fig:A0fNL}, we plot the distribution of $A_0$, using Eq.(\ref{eq:monopolemodulation}), for  $\fNL=0, 50, 100$, along with the distribution obtained from the numerically generated Sachs-Wolfe CMB maps.

\begin{figure*}
 \includegraphics[width=0.48\textwidth]{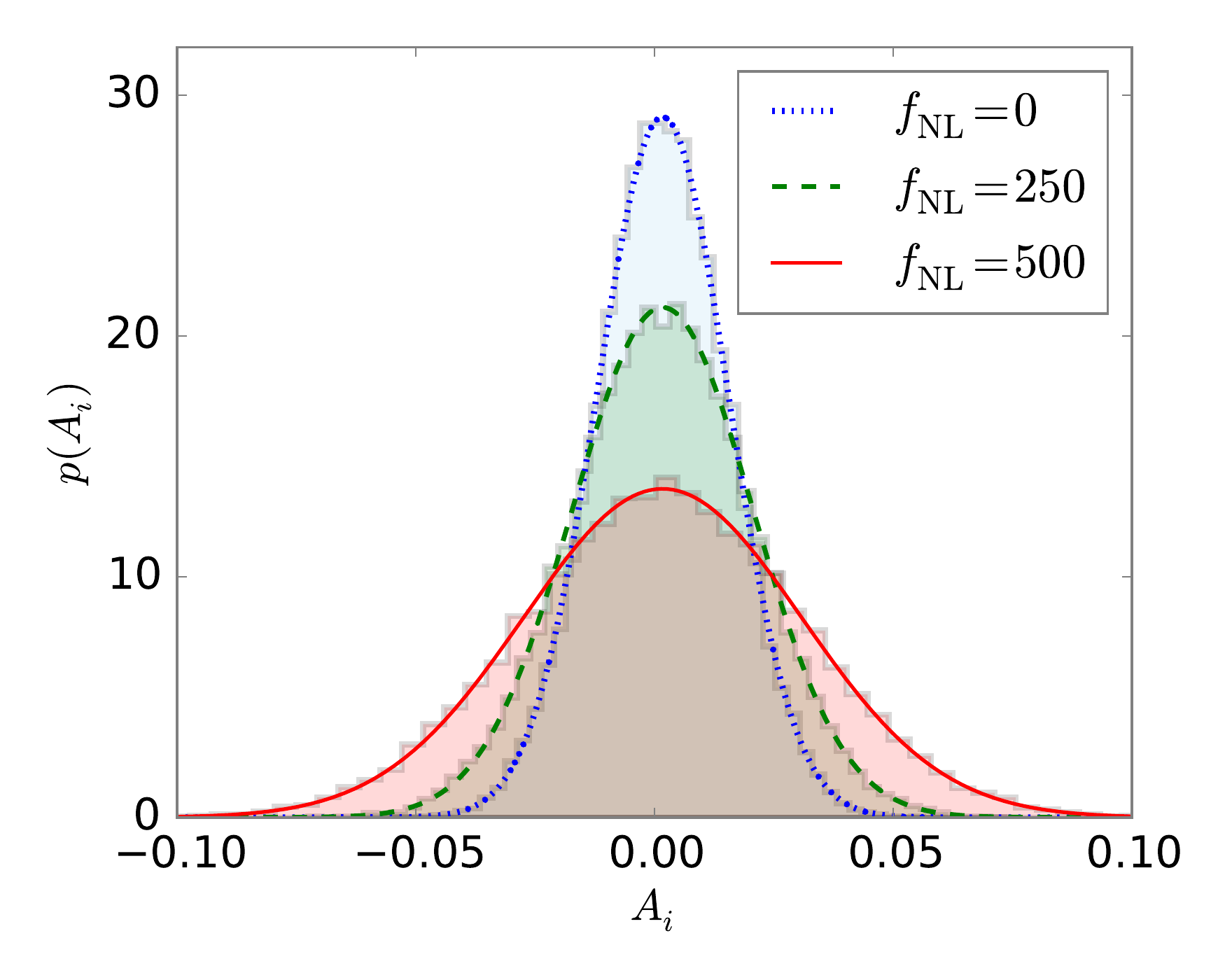}
  \includegraphics[width=0.48\textwidth]{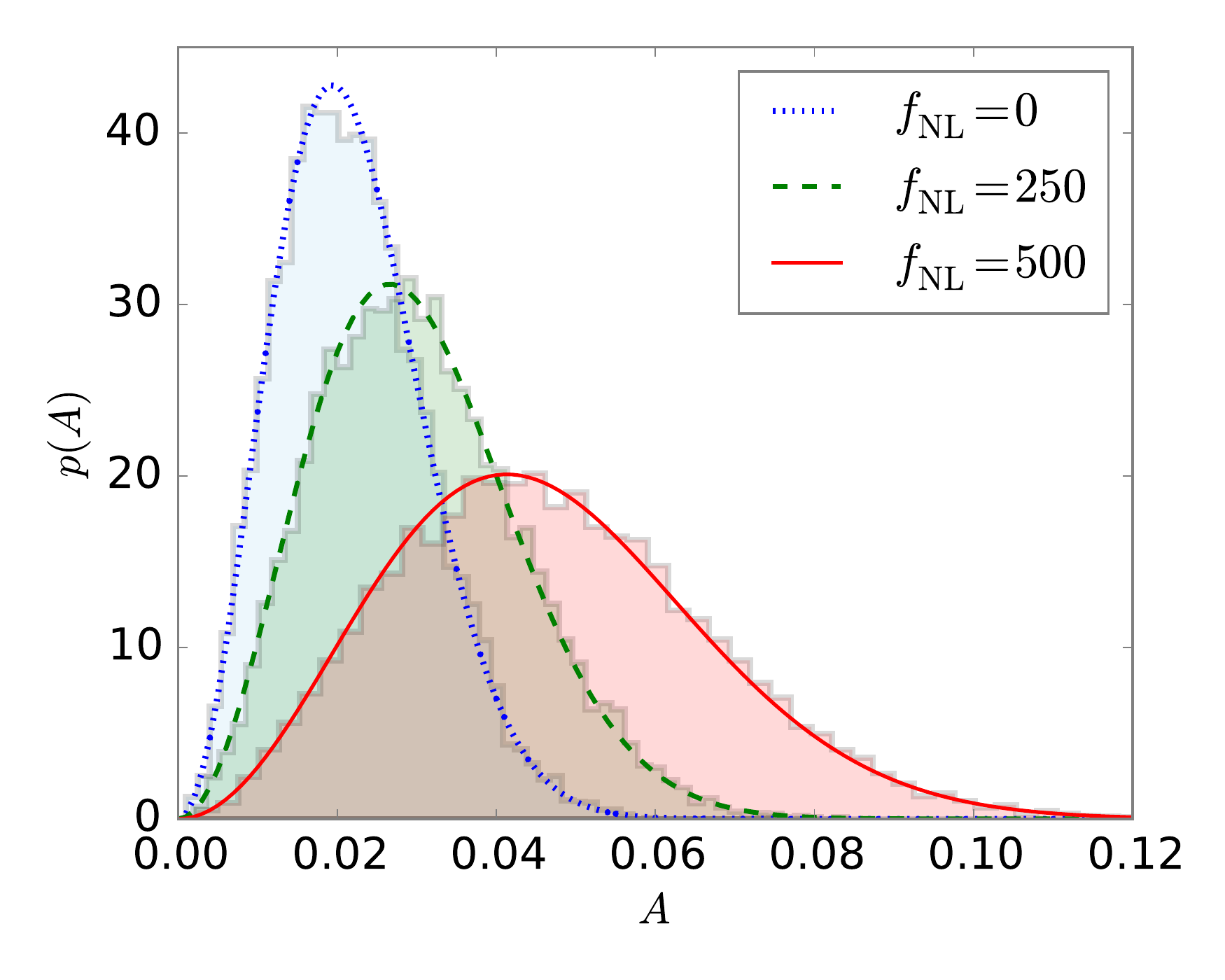}
 \caption{(color online). \textit{Left}: The distribution of power asymmetry $A_i$ (in a particular direction $d_i$) measured in 10000 simulated CMB skies as described in the text. From each simulated map, three $A_i$ values are generated in three orthonormal directions in the sky. The dotted blue line for the Gaussian CMB maps is the best-fit normal distribution curve; this gives us the Gaussian cosmic variance standard deviation $\sigma_G$. The curves (dashed green and solid red) for the non-Gaussian models are normal distributions with zero mean and variance given by $\sigma^2=\sigma_G^2 + \sigma_{\fNL}^2$, where $\sigma_{\fNL}$ is computed using Eq.(\ref{eq:sigmafNL}). \textit{Right}: The distribution of  the amplitude of power asymmetry $A$ from the simulated Gaussian and non-Gaussian CMB maps. The lines are the computed Maxwell distributions for corresponding $\sigma$ values and match the distributions obtained in numerical realizations very well. Note that the figures above were generated for the single source local model. If we used the mixed inflaton-curvaton two-field extension with the curvaton power fraction $\xi<1$, we would get the above distributions for smaller values of $\fNL$. For example, for $\xi=0.25$, the distributions shown for $\fNL=500$ above would be generated by a smaller $\fNL=\sqrt{0.25}\times 500=250$.}
 \label{fig:Adist}
\end{figure*}

\subsection{Dipole modulation ($L=1$)}
A dipole modulation of the power spectrum defined as in Eq.(\ref{eq:PPhidipole}) generates a hemispherical power asymmetry with the same amplitude $2A_i$. Therefore, we will look at the quantity:
\be
 A_i=\frac{1}{\sum_{\ell=2}^{\ell_{\rm max}} (2\ell+1)}\sum_{\ell=2}^{\ell_{\rm max}} (2\ell+1) \frac{\Delta \Cl}{2 \Cl},
 \label{eq:A}
\ee
with $\ell_{\rm max}=100$ and $\Delta \Cl = \Cl^+ - \Cl^-$, where $+$ and $-$ refer to two hemispheres in some direction $d_i$. We will consider $A_i$s in three orthonormal directions $d_1$, $d_2$, and $d_3$ on the sky. Each $A_i$ in a particular direction $d_i$ is normally distributed with zero mean. The variance $\sigma_G^2$ can be measured from the numerical realizations of Gaussian Sachs-Wolfe CMB maps and depends on the CMB multipoles used in Eq.(\ref{eq:A}) and the value of the $\Cl$s. For non-Gaussian maps, the distribution of the $A_i$s have an increased variance given by: $\sigma^2 = \sigma_{\fNL}^2 + \sigma_G^2$, where $\sigma_{\fNL}$ is given by Eq.(\ref{eq:sigmafNL}). The power asymmetry dipole amplitude for each CMB sky is then $A=\left(A_1^2+A_2^2+A_3^2\right)^{\frac{1}{2}}$. The probability distribution function (pdf) of $A$ is the $\chi$ distribution (or the Maxwell distribution):
\be
 p_\chi(A, \sigma) = \sqrt{\frac{2}{\pi}} \frac{A^2}{\sigma^3} {\rm Exp}\left[\frac{-A^2}{2 \sigma^2}\right],
 \label{eq:pchi}
\ee
where $\sigma = \sqrt{\sigma_{\fNL}^2 + \sigma_G^2}$. Figure \ref{fig:Adist} shows that the distribution of asymmetry amplitudes obtained from the CMB realizations agree extremely well with the $\chi$ distribution given above. Note that only $\sigma_G$ is measured from the numerical maps; $\sigma_{\fNL}$ is directly computed for a value of $\fNL$ using Eq.(\ref{eq:sigmafNL}).

\subsection{Statistical analysis}
\label{sec:stat}
In this section, we present examples of how we can perform a statistical analysis using the results from the previous section for the distributions of power modulation on the Sachs-Wolfe CMB sky. While direct comparison of the amplitudes obtained in our Sachs-Wolfe CMB realizations with the reported values of power asymmetry $A$ is not possible, we can make a connection between our simpler case and the asymmetry in the observed CMB sky by using the $p$-value of the asymmetry. For a given measurement of $A$ and the normalized pdf for $A$, $p(A)$ [which in our model depends on $\fNL$, see Eq.(\ref{eq:pchi})], the $p$-value is simply given by $\int_A^{\infty} p(A') dA'$; i.e. it gives the probability that the observed value of the asymmetry amplitude is greater than some threshold value $A$. We find that an asymmetry amplitude of $A=0.055$ is approximately $3.3 \sigma$, i.e. a $p$-value of $0.001$ with respect to the distribution of $A$ obtained in our Gaussian Sachs-Wolfe CMB maps. This is approximately equal to some of the more recent reports for the significance of the hemispherical power asymmetry \cite{Akrami:2014eta, Ade:2015sjc}. Therefore, we will use $A=0.055$ as the value of the asymmetry when making connections with the observations of the anomaly.

\begin{figure}
 \includegraphics[width=0.48\textwidth]{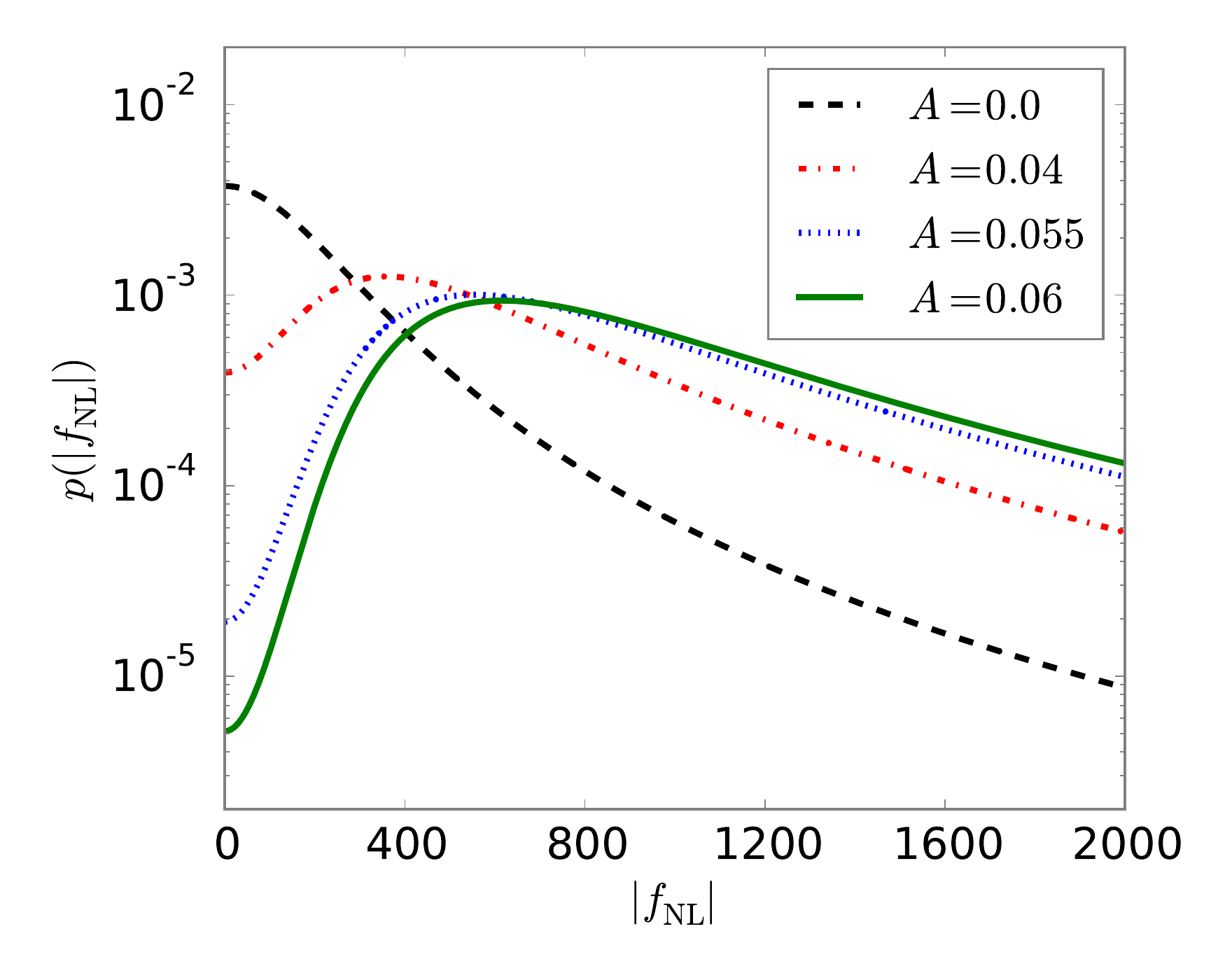}
 \caption{(color online). The posterior probability distribution of $|\fNL|$ values for different observed amplitudes $A$ of power asymmetry. Only large-scale CMB modes $l\leq 100$ are used to compute the cosmic variance pdf for $A$. $A=0.055$ corresponds to a $p$-value of $0.001$ in our $\fNL=0$ numerical maps. This is about $3.3 \sigma$, approximately equal to some of the reported significance of the power asymmetry anomaly \cite{Akrami:2014eta, Ade:2015sjc}. Although our formula for the expected asymmetry becomes less accurate for larger values of $\fNL$ we have checked that it approximates the numerical results quite well even for $\fNL=2000$. Therefore, the shape of the posterior distributions obtained above in the $\fNL$ window shown will not be affected by the inaccuracy of the formula at larger $\fNL$ values. However, the change in the distribution for larger $\fNL$ can change the normalization.}
 \label{fig:posteriorA1}
\end{figure}

\begin{figure}
 \includegraphics[width=0.48\textwidth]{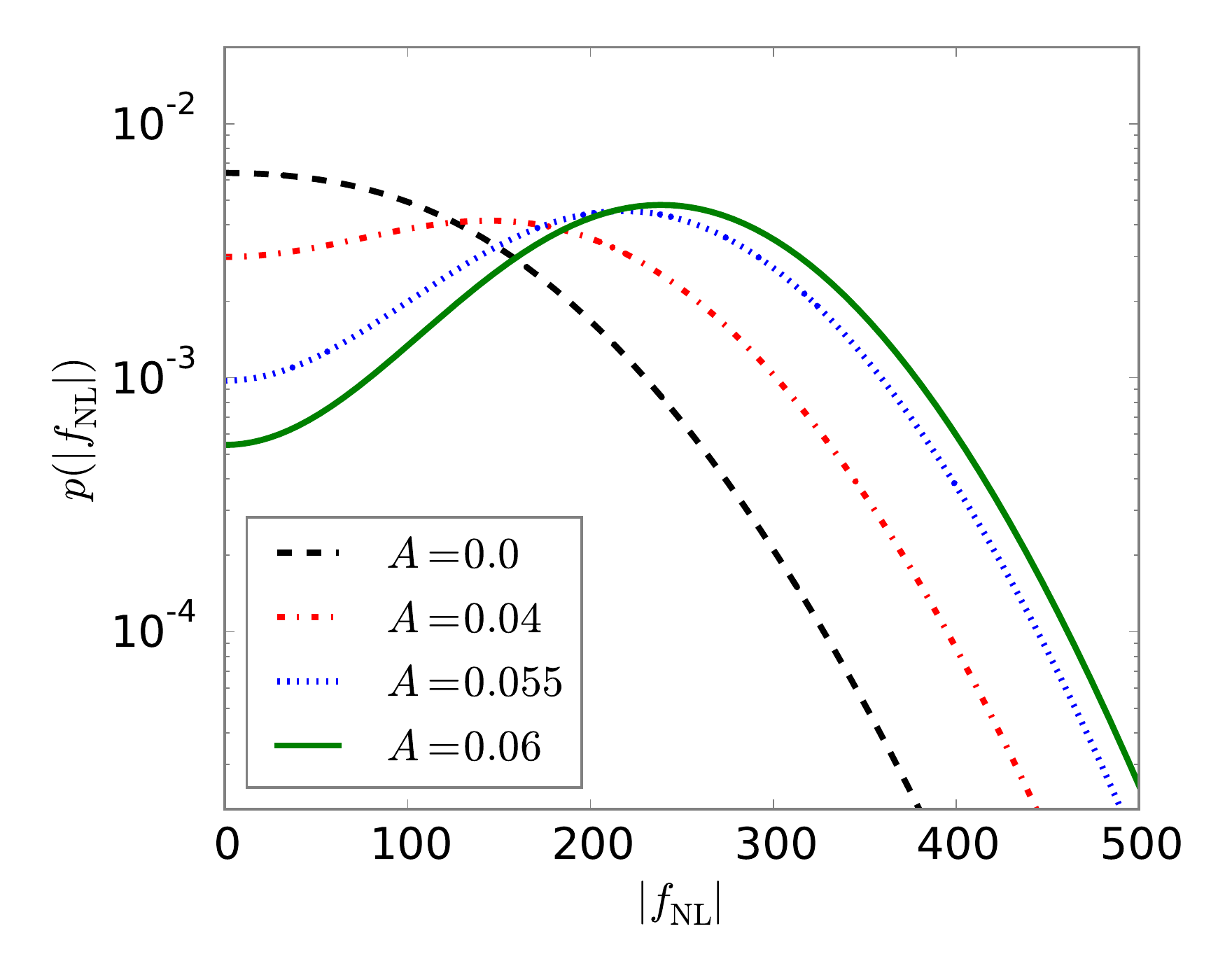}
 \caption{(color online). The posterior probability distribution of $|\fNL|$ after combining the bispectrum constraints at large scales ($l\lesssim 100$, $\fNL=-100\pm100$) with the power asymmetry constraint for the given value of $A$.}
 \label{fig:posteriorfNLcombined}
\end{figure}

\begin{figure}
 \includegraphics[width=0.48\textwidth]{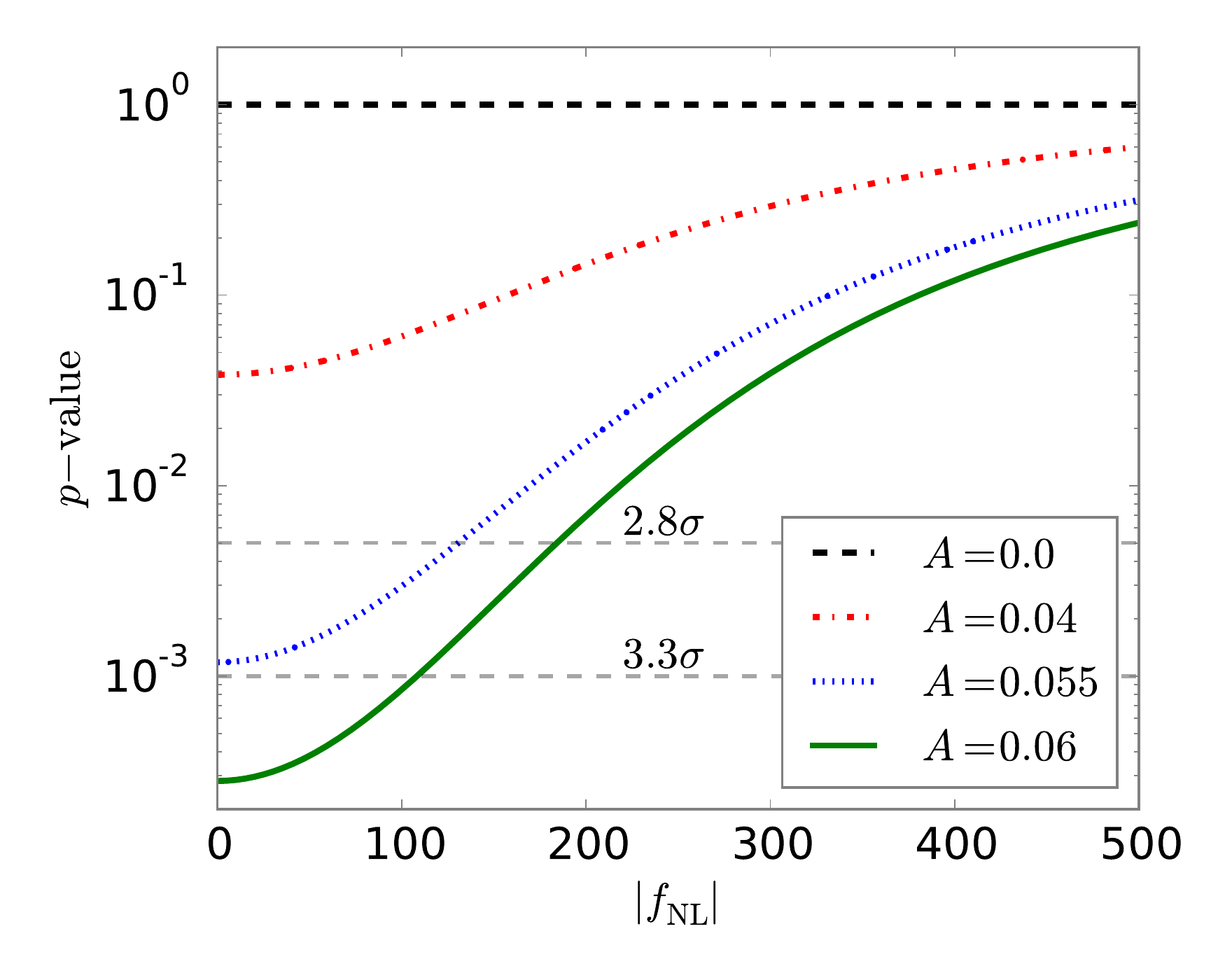}
  \caption{(color online). The $p$-value for different values of asymmetry amplitudes $A$ (i.e. the probability of obtaining an asymmetry amplitude equal to or greater than $A$) in a local non-Gaussian model as a function of the value of $|\fNL|$. For $A=0.055$, we see that the significance goes below  $3\sigma$ around $|\fNL| \approx 100$. For the two-field extension of the local model (with a curvaton power fraction $\xi$) described in the text, the x-axis should be labeled $|\fNL|/\sqrt{\xi}$. Then, for $\xi<1$, a smaller bispectrum amplitude $|\fNL|$ (by a factor of $\sqrt{\xi}$ than labeled in the figure above) is required to achieve a $p-$value shown in the figure above.}
 \label{fig:pvalues}
\end{figure}

When we have a measurement of the power asymmetry amplitude $A$, we can write the likelihood for $\fNL$ as $\mathcal{L}(\fNL|A) = p_\chi(A, \sigma) \nonumber$, whose expression is given in Eq.(\ref{eq:pchi}). From this likelihood, we can infer the posterior distribution for $\fNL$ given a measurement of $A$. We can interpret the statistics in different ways:

\begin{itemize}
 \item We can use any power asymmetry as a signal of local non-Gaussianity.  Using only the large-scale CMB multipoles ($l\leq100$), for a given value of $A$, we can obtain the posterior distribution for $|\fNL|$ (averaged) for the corresponding range of scales. In Figure \ref{fig:posteriorA1}, we plot the $\fNL$ posterior for a few values of the asymmetry $A$, assuming a uniform prior on $|\fNL|$.  
 
 \item We can combine the large-scale bispectrum constraints on $\fNL$ with the constraints from the power asymmetry $A$. For this, we use a rough estimate of $\fNL$ for $\ell \lesssim 100$ of $\fNL=-100\pm100 (1\sigma)$ (estimated from Figure 2 of \cite{Smith:2009jr}). We assume that the $\fNL$ posterior from WMAP is a normal distribution. However, since the power asymmetry is only sensitive to the magnitude of $\fNL$ and not the sign, we use the folded normal distribution (given $\fNL^{\rm measured} = \mu \pm \sigma$):
\begin{align*}
 p_{\rm fold}(\fNL) =
  \begin{cases}
 \, \frac{1}{\sigma\sqrt{2\pi}} \left[e^{-\frac{1}{2}\left(\frac{\fNL-\mu}{\sigma}\right)^2}\right.& \\
  \;\;\;\;\;\left.+\, e^{-\frac{1}{2}\left(\frac{\fNL+\mu}{\sigma}\right)^2} \right] & \fNL\geq0\\
 \,0 &\fNL<0.
  \end{cases}
 \label{eq:fold}
\end{align*}
Then, we multiply the above pdf with $\mathcal{L}(\fNL|A)$ to get the combined likelihood from which we can get the posterior for $\fNL$ after normalizing. We show the combined posterior distribution of $\fNL$ for a few power asymmetry amplitudes $A$ in Figure \ref{fig:posteriorfNLcombined}.

\item 
Although the power asymmetry data alone ($A\approx 0.055$) prefers $\abs\fNL \approx 500$ as the most likely value (or $\abs\fNL\approx 200$ when the bispectrum constraints are also applied), the probability of an asymmetry increases whenever $\fNL\neq0$. In Figure \ref{fig:pvalues} we quantify how the probability of an observed dipole modulation changes as $|\fNL|$ increases. For example, the $p$-value of $0.001$ for $A=0.055$ changes to $0.046$ for $|\fNL|=265$ (which is within the $2\sigma$ window of the large-scale bispectrum constraint). In other words, even an amplitude of non-Gaussianity well below $\abs\fNL \approx 500$ renders the observed asymmetry less ``anomalous." 

\end{itemize}

\subsection{Bayesian evidence}
The previous section demonstrated that the power asymmetry data can be used to constrain non-Gaussian models, and that the amplitude of the observed asymmetry is less ``anomalous" when non-Gaussianity is included. However, we also need to ask whether the data are such that the non-Gaussian model is preferred over the Gaussian.

To compare the posterior odds for different models $\mathcal{M}_i$, given the data $\vec{y}$, we compute the Bayes factor 
\be
B_{12}=\frac{p(\vec{y}\,|\mathcal{M}_1)}{p(\vec{y\,}|\mathcal{M}_2)},
\ee
where the factors in the numerator and denominator are the model likelihoods for models 1 and 2 respectively. In the simplest comparison, we take $\fNL$ as the only parameter of the models. The data we consider include the measured amplitude of the power asymmetry and the CMB constraint on the amplitude of the local bispectrum on large angular scales. For an introduction to Bayesian statistical methods applied to cosmology, see for example \cite{Trotta:2008qt}.

The non-Gaussian model reduces to the isotropic Gaussian model for $\fNL=0$ (while the probability of the power asymmetry remains non-zero). In that case the evaluation of the Bayes factor can be simplified and a direct Bayesian model comparison can be done using the Savage-Dickey density ratio (SDDR) \cite{Trotta:2005ar, Trotta:2008qt}. The SDDR is given by,
\be
 B_{01} = \frac{p(\theta_i | \vec{y}, \mathcal{M}_1)}{\pi(\theta_i|\mathcal{M}_1)}\bigg|_{\theta_i=\theta_i^*}
\ee
where $\mathcal{M}_1$ is the more complex model (non-Gaussian in our case) that reduces to the simpler model $\mathcal{M}_0$ (Gaussian) when the set of parameters $\theta_i$ goes to $\theta_i^*$ ($\fNL\rightarrow\fNL=0$). Here,  $p(\theta_i | \vec{y}, \mathcal{M}_1)$ is the posterior for $\fNL$ (plotted in Figure \ref{fig:posteriorfNLcombined}) and $\pi(\theta_i|\mathcal{M}_1)$ represents the prior for the parameter in the complex model $\mathcal{M}_1$. Our current case only has one parameter ($\fNL$) and one datum (the dipolar asymmetry $A$). While there may be other interesting possibilities to consider for the prior probability of $\fNL$, we illustrate the calculation of $B_{01}$ above using the constraint on the parameter $\fNL$ from large-scale bispectrum measurements as reported by the WMAP and Planck missions as the prior.

In Table \ref{table:sddr}, we list SDDR for a few values of $A_{\rm obs}$. For the prior, we have used the folded normal distribution for $\fNL=-100\pm100$ which is a rough estimate of $\fNL$ for the largest scale i.e. up to $\ell=100$  from \cite{Smith:2009jr}. Note that the only value from the prior pdf that is used to compute the SDDR is $\fNL=0$, so the above consideration from large-scale $\fNL$ constraints is the same as using a uniform prior for $|\fNL|$ in the range $\left(0, \frac{1}{\pi(\fNL=0)}\right) \approx \left(0, 207\right)$. If the prior range is expanded, then the magnitude of $B_{01}$ increases thereby reducing the evidence for non-zero $\fNL$. For $A_{\rm obs}=0.055$ (whose $p$-value roughly corresponds to observed $A$), the strength of evidence for a non-zero $\fNL$ is between weak and moderate in the empirical (Jefferys') scale \cite{Jeffreys} quoted, for example, in Table 1 of \cite{Trotta:2008qt}.

\begin{table}[htbp]
  \begin{center}
   \begin{tabular}{|r|c|c|c|}
   \hline  $A_{\rm obs}$ & $p$-value & SDDR ($B_{01}$) & $\ln{B_{01}}$
\\     \hline \hline
	0.02  & 0.5511  & 1.1362 & 0.128 \\ \hline
	0.04  & 0.0381  & 0.6174 & -0.482 \\ \hline
	0.05  & 0.0043  & 0.3211 & -1.136 \\ \hline
	0.055 & 0.001  & 0.2012 & -1.603  \\ \hline 
	0.06  & 0.0003  & 0.1123 & -2.186 \\ \hline
\hline
  \end{tabular}
   \end{center}
\caption{Savage-Dickey Density Ratio (SDDR) for different observed values of dipole power modulations at large scales. The $p-$values listed above are computed with respect to the distribution of $A$ values from Gaussian CMB maps. A reasonable value to compare to various reports of the observed hemispherical power asymmetry is a $p-$value of 0.001 ($\approx 3.3 \sigma$) i.e. $A=0.055$. Values of $|\ln{B_{01}}|=5.0, 2.5$ suggest strong and moderate evidence, respectively \cite{Trotta:2008qt}. In our convention above, a negative value for the logarithm of the Bayes factor means the evidence is in favor of the more complex non-Gaussian model $\mathcal{M}_1$.}
\label{table:sddr}
 \end{table}

The results in Table I show that the data we have used, at least in this simple analysis, show no more than a weak preference for the non-Gaussian model. A more thorough analysis is unlikely to change this conclusion very much: in \cite{Gordon:2007xm}, the authors use earlier studies of the power asymmetries in the WMAP data to put the best possible Bayesian evidence of $\ln{B_{01}} \approx -2.16$ corresponding to odds ($\leq 9:1$, weak support). The method to compute the maximum possible Bayesian evidence is based on Bayesian calibrated $p$-values \cite{Sellke}. The $p-$value used from the data analysis of the 3-year WMAP maps was $p=0.01$ \cite{Eriksen:2007pc}. If one instead used $p=0.001$, which is approximately the level of significance from various more recent analyses of Planck and WMAP temperature anisotropy maps, the best possible Bayesian evidence in favor of the $A\neq0$ anisotropic model becomes $\ln{B}_{01} \approx -4.0$ corresponding to the odds ($\leq 50:1$). A $p-$value of 0.0003 (about $3.6\sigma$) is necessary to obtain a best possible $\ln{B_{01}} \approx -5.0$, which implies strong evidence \cite{Gordon:2007xm}. 

\section{Beyond the local ansatz}
\label{sec:goodModel}
The previous section discussed in detail the effect of local-type non-Gaussianity with constant $\fNL$ that couples a gradient (induced by superhorizon modes) across the CMB sky to the observable modes. We demonstrated that a dipolar asymmetry is expected in models with local-type non-Gaussianity. Of course, local non-Gaussianity as the source of the asymmetry is only compatible with the data if we restrict ourselves to the largest scales. Both the amplitude of the asymmetry and the amplitude of non-Gaussianity must sharply decrease on smaller scales. The question then is whether there is a different model of non-Gaussianity that is consistent with all observational constraints and generates the observed asymmetry in detail. If so, does the current data favor this model over the isotropic, Gaussian assumption? Could future data ever favor such a model?

To address some of these questions, we will first demonstrate that scale-dependent modulations are a generic feature of non-Gaussian models other than the local model. We will then construct a scenario that is more likely to be preferred by the data by considering scale-dependent local non-Gaussianity. Finally, we will  provide examples of evaluating the Bayesian evidence for this scenario. Although a model of non-Gaussianity beyond the local ansatz may add more parameters, if the model has other consequences in the data we might hope to find more evidence for it. This is particularly true if the model has measurable effects on smaller scales, where the usual cosmic variance for Gaussian models is smaller.

\subsection{Power asymmetry from general bispectra}
We can easily extend the inhomogeneous power spectrum calculation in the presence of local non-Gaussianity to other bispectrum shapes. For example, consider that a Fourier mode of the Bardeen potential is given by \cite{2015PhRvD..91h3518B}:
\begin{eqnarray}
 \Phi(\kv) = \phi(\kv) + \frac{\fNL}{2} \int \frac{d^3\qv_1}{\tpc} \int d^3\qv_2\; \phi(\qv_1) \phi(\qv_2) \nonumber \\ N_2(\qv_1, \qv_2, \kv) \diracdelta(\kv-\qv_1-\qv_2) +\dots
 \label{eq:generalNG}
\end{eqnarray}
where as before $\phi$ is a Gaussian field. The kernel $N_2$ can be chosen to generate any desired bispectrum and the dots represent terms higher order in powers of $\phi$ (which generate tree-level $n$-point correlations). Considering only the generic quadratic term, the power spectrum in sub-volumes can be computed as in the case of the local bispectrum (see Appendix \ref{app:derivation}), and we get $\PPhiS$:
\be
 P_\phi(k) \left[1 + 2\fNL \int \frac{d^3\kv_{\ell}}{\tpc} \phi(\kv_{\ell}) N_2(\kv_{\ell}, -\kv, \kv) e^{i\kv_{\ell}\cdot \xv}\right].
 \label{eq:modulatedpower}
\ee
From the form of the above equation, one can see that a $k$-dependent power modulation is a feature of non-local non-Gaussianity i.e. the $k$ dependence of the kernel $N_2$ is carried by the modulated component of the power spectrum in the small volume. The kernels for local, equilateral and orthogonal bispectrum templates are \cite{2015PhRvD..91h3518B}:
\begin{eqnarray}
 N_2^{\rm local} &=& 2; \nonumber\\
 N_2^{\rm ortho} &=& \frac{4 k_{\ell}^2 -2 kk_{\ell}}{k^2};\nonumber \\
 N_2^{\rm equil} &=& \frac{2k_{\ell}^2}{k^2}.
 \label{eq:N2kernels}
\end{eqnarray}

\begin{figure}
 \includegraphics[width=0.48\textwidth]{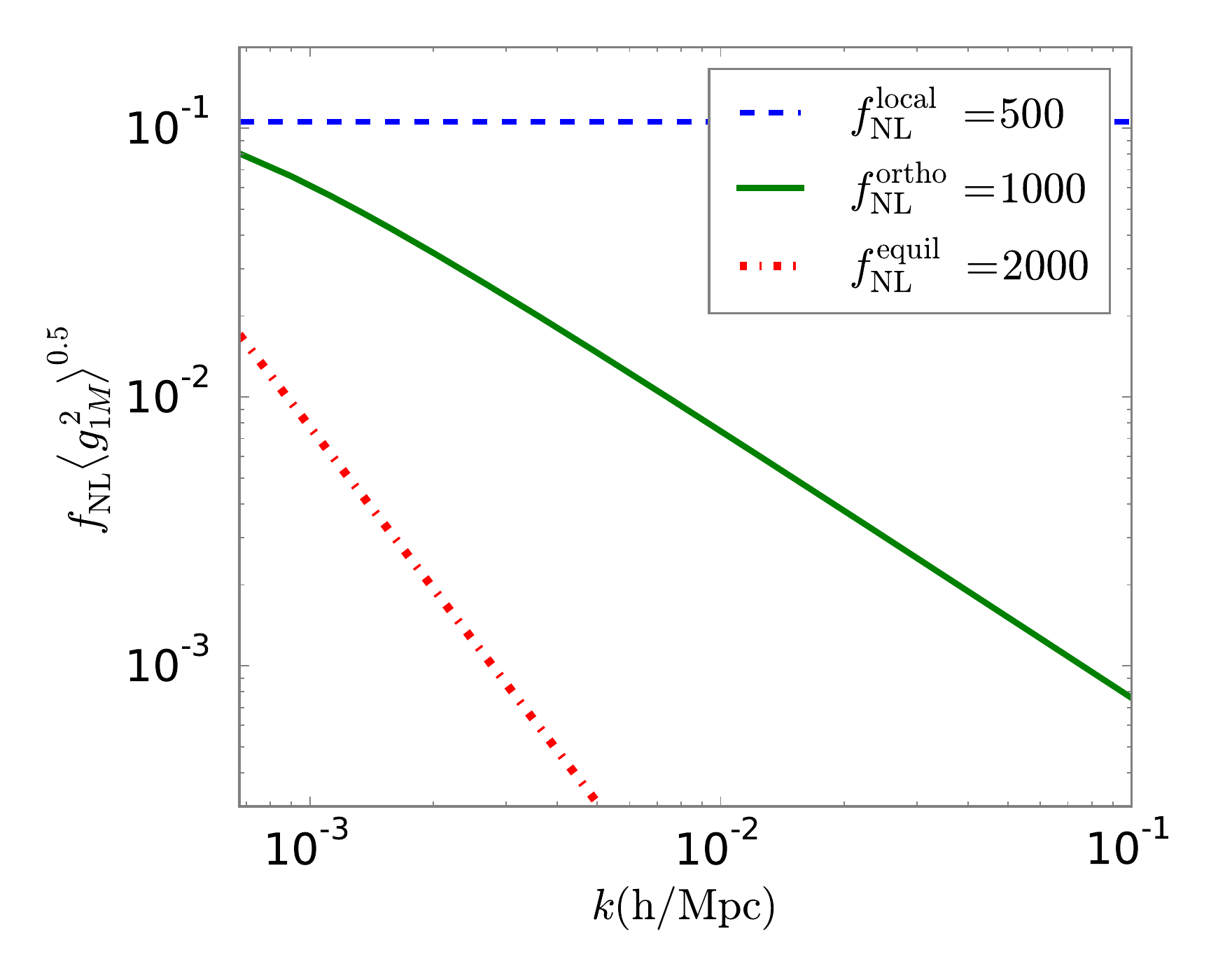}
 \caption{(color online). The expected dipolar modulation of the power spectrum for large amplitude local-, orthogonal- and equilateral-type non-Gaussianities. While the amplitude of expected modulation is smaller for non-local shaped bispectra, the modulation generated by them is scale dependent.}
 \label{fig:dipole}
\end{figure}

\begin{figure}
 \includegraphics[width=0.48\textwidth]{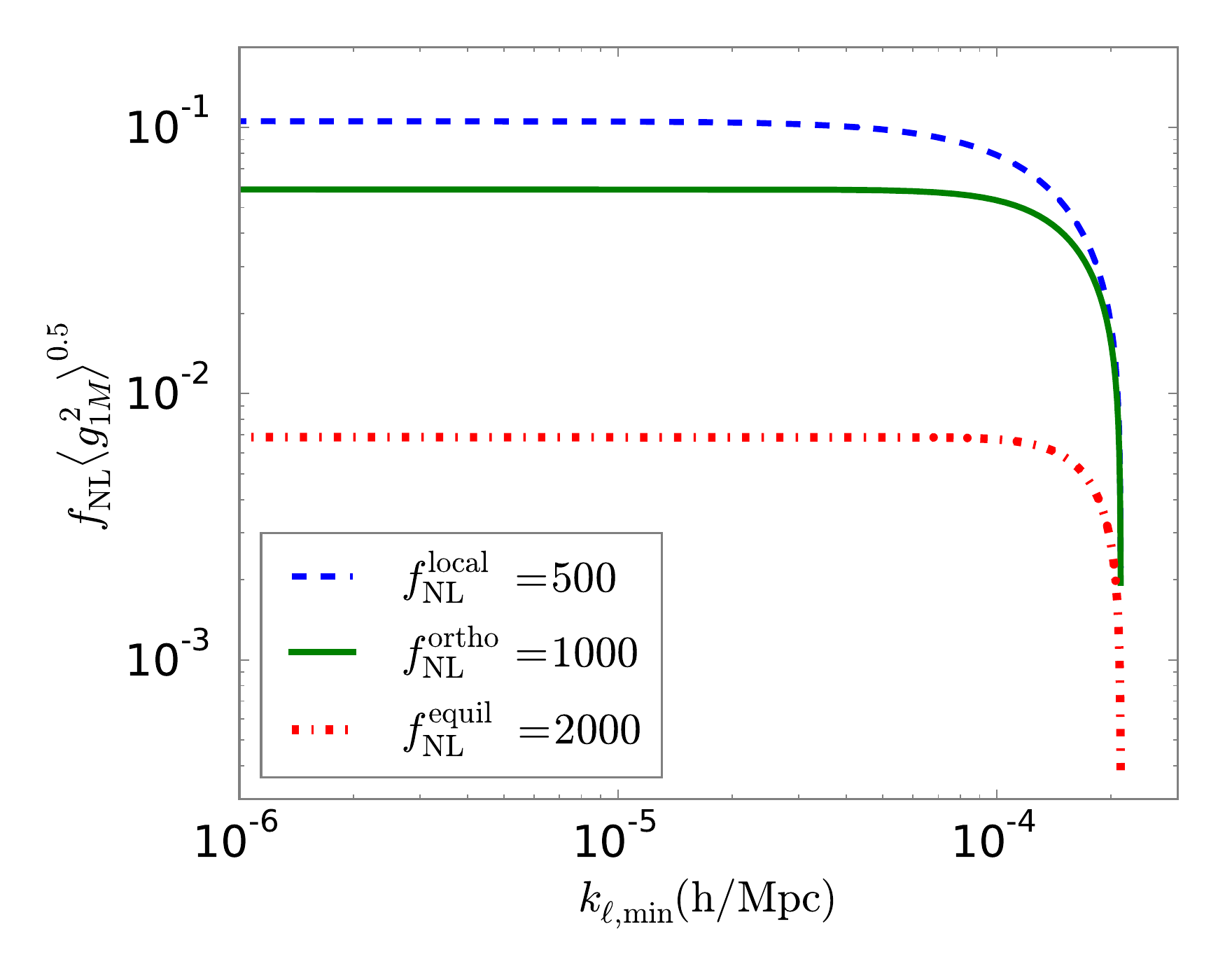}
  \caption{(color online). The expected amplitude of the power asymmetry from superhorizon modes ($k_{\ell} < \pi/\rcmb$), as a function of the minimum wavenumber $k_{\ell, {\rm min}}$ considered to compute $\langle \g_{1M}^2 \rangle^{0.5}$, for large-amplitude local-, orthogonal- and equilateral-type non-Gaussianities. We can see that most of the contribution is due to the modes with wavelengths almost the size of the observable universe i.e. within one e-fold. We get similar behaviors for the monopole modulations (not shown in the figure) except for the case of local non-Gaussianity. For local non-Gaussianity, the monopole modulation gets contributions from arbitrarily small $k$ modes and $\langle g_{00}^2 \rangle_{\rm local}$ becomes infrared divergent.}
 \label{fig:supermodes}
\end{figure}

If one uses the kernel for equilateral- or orthogonal-type non-Gaussianities, then the monopole shifts are not infrared divergent. However, the magnitudes of modulation (both monopole and dipole) are smaller compared to the local case i.e. a very large amplitude of $\fNL^{\rm equil}$ or $\fNL^{\rm ortho}$ is necessary for the effect to be interesting. For example, we plot the expected modulation amplitude for local-, equilateral- and orthogonal-type non-Gaussianities in Figure \ref{fig:dipole}. In Figure \ref{fig:supermodes}, we illustrate that for local, orthogonal and equilateral bispectra, the power asymmetry is generated by perturbation modes that lie just outside the horizon. The quasi-single field model \cite{Chen:2009zp} may also be interesting to consider: it has a scale-independent bispectrum with a kernel that varies between the local and equilateral cases depending on the mass of an additional scalar field coupled to the inflaton. 

In general then, if the power asymmetry is coming from mode coupling, the fact that the observed asymmetry falls off on small scales implies that shorter scales are more weakly coupled to superhorizon modes than larger scales are. This is possible with either a scale-independent bispectrum (as the equilateral and orthogonal cases above demonstrate) or with a scale-dependent bispectrum.

\subsection{Generating a scale-dependent power asymmetry}
To match the observed scale dependence of the power asymmetry anomaly, the strength of coupling of subhorizon modes to the long wavelength background must be scale dependent. The relevant scale dependence in this context can be fully parametrized by introducing two bispectral indices that capture the scale dependence in our observable volume and a more general coupling strength to the long wavelength modes: 
\begin{eqnarray}
\label{eq:twoparammodulation}
P_{\Phi,S}(k,\xv) &=& P_{\phi}(k)\left[1+4\fNL(k_0)\left(\frac{k}{k_0}\right)^{n_f}\right.\\\nonumber
&&\left.\times\int \frac{d^3 \kv_{\ell}}{\tpc}\left(\frac{k_{\ell}}{k_0}\right)^{\alpha} \phi(\kv_{\ell}) e^{i\kv_{\ell} \cdot \xv}\right]\;.
\end{eqnarray}
Here $n_f<0$ turns off any power asymmetries on shorter scales. The parameter $\alpha<0$ enhances the sensitivity of the model to infrared modes (as used in \cite{Schmidt:2012ky, Agullo:2015aba}). In the case $\alpha\leq-1$, the dipole asymmetry would be infrared divergent in a universe with a scale-invariant or red-tilt power spectrum. Notice that scale-invariant bispectra always have $n_f=-\alpha$, so any scale-invariant bispectrum that increases IR sensitivity also increases the expected asymmetry on smaller scales. Finally, although we have used $\fNL$ to label the coefficient above, a similar expression can be derived from higher-order correlation functions (e.g., to capture the effects of $\gNL$ \cite{Kenton:2015jga}). Previous discussions of the power asymmetry from scale-dependent non-Gaussianity include \cite{Erickcek:2009at, Lyth:2013vha, Kohri:2013kqa, Lyth:2014mga, Firouzjahi:2014mwa, Byrnes:2015asa, Agullo:2015aba}.

In principle, additional data could eventually constrain all of the parameters introduced above (or at least their values on subhorizon scales). However, here we will consider only one additional measurement (the large-scale power suppression) and so we will restrict our attention to the case with just one additional parameter. We take a local-shape bispectrum with an amplitude that depends on the scale of the short wavelength mode as
\be
 \fNL(k)= \fNL^0\left(\frac{k}{k_0} \right)^{n_{\fNL}}.
 \label{eq:fnlofk}
\ee
In terms of the parameters in Eq.(\ref{eq:twoparammodulation}), $n_f=n_{\fNL}$ and $\alpha=0$.

In addition to a scale-dependent power asymmetry and a scale-dependent bispectrum amplitude, a scale-dependent local non-Gaussianity also generates a scale-dependent modulation of the power-spectrum amplitude. This is the monopole power modulation ($L=0$) discussed in Section \ref{sec:cmb} (a similar point was made in \cite{Lyth:2014mga}). When local-type non-Gaussianity has a scale-independent amplitude, the power spectrum amplitude is modulated similarly at all scales, thereby making the effect unobservable.  However, the scale-dependent case is more interesting as it makes the power modulation scale dependent and therefore an observable effect. This can be easily seen by generalizing Eq.(\ref{eq:monopole}) for the case of $\fNL(k)$:
\be
 P_\Phi(k) = P_\phi(k) \left[ 1 + \fNL(k) \frac{g_{00}}{2\sqrt{\pi}}\right].
\ee
As previously discussed, $g_{00}$ is normally distributed with zero mean and the variance $\langle g_{00}^2 \rangle$ requires a cutoff to limit contributions from arbitrarily large modes, which we have earlier parametrized as the number of superhorizon efolds of inflation. While the above formula is only valid for small modulations $A_0$, Figure \ref{fig:A0fNL} shows that the formula is quite accurate at least up to $|A_0|\approx0.3$. There is an additional subtlety because,  in the presence of a monopole shift, the observed bispectrum on large scales will not be exactly of the form in Eq.(\ref{eq:fnlofk}).  However, the difference is small for small $A_0$.

\begin{figure}
 \includegraphics[width=0.48\textwidth]{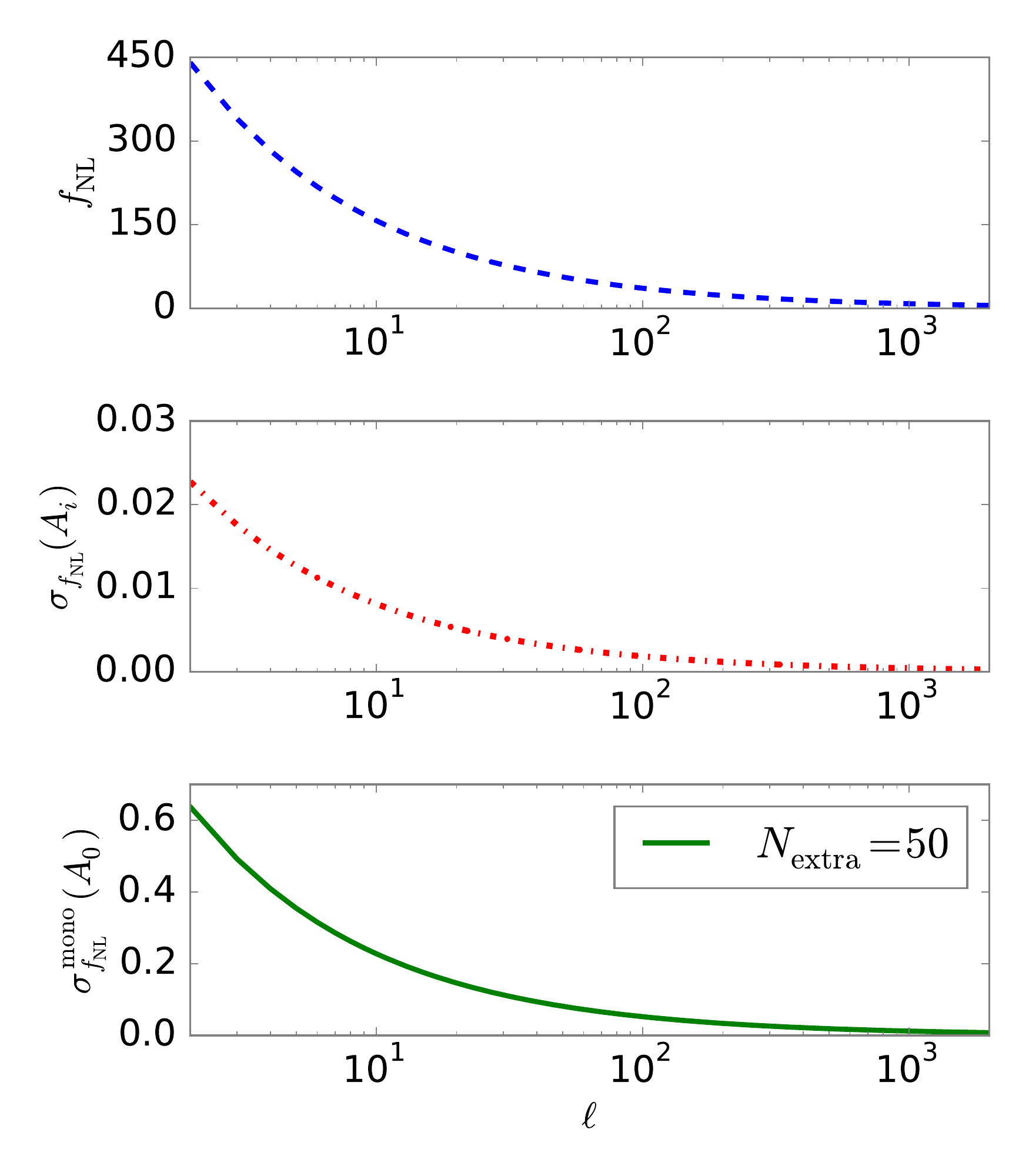}
 \caption{(color online). The effect of a simple scale-dependent local non-Gaussian model, $\fNL(\ell)=50(\ell/60)^{-0.64}$, in the bispectrum (top), the power asymmetry amplitude (middle), and the modulation of power spectrum amplitude (bottom). A different choice of $\fNL^0$ will only rescale the lines above.}
 \label{fig:scaledepfNL}
\end{figure}

Consider a simple example of scale-dependent local non-Gaussianity given by $n_{\fNL}=-0.64$, $k_0=60 (\pi/\rcmb)$ and $\fNL^0 = 50$. In multipole space, one can approximate $\fNL^0 = \fNL(\ell=60) = 50$ and $\fNL(\ell) = 50 (\ell/60)^{-0.64}$. These numbers are chosen to facilitate comparison with the scale-dependent modulation model results in \cite{Aiola:2015rqa} (see Table I therein). In Figure \ref{fig:scaledepfNL}, we plot $\fNL$, $\sigma_{\fNL}$ (the expected $1\sigma$ amplitude of the power asymmetry in a particular direction due to $\fNL$), and $\sigma_{\fNL}^{\rm mono}$ (the expected $1\sigma$ shift in the amplitude of the power spectrum due to $\fNL$), as a function of the multipole number $\ell$. The purpose of the figure is to illustrate how  scale-dependent local non-Gaussianity can produce more than one signature in the CMB. Therefore, we have not included the effect of the Gaussian cosmic variance, which would add variance to both the power asymmetry and monopole modulation amplitudes; this becomes important when performing parameter estimation of $\fNL^0$ and $n_{\fNL}$. While such a full parameter estimation analysis is beyond the scope of this work, we illustrate in the next section that, in a simplified context, adding the $A_0$ constraint can be useful in some situations.

Notice that the scale-dependent non-Gaussian model of Eq.(\ref{eq:fnlofk}) also generates asymmetries in the spectral index:
\be
\left.\frac{d{\rm ln} P}{d{\rm ln}k}\right|_{k_0} = (n_s-1)_{\phi}+n_{\fNL}\frac{\Delta P(k_0,\xv)}{1+\Delta P(k_0,\xv)},
\ee
where $(n_s-1)_{\phi}$ is the spectral index of the Gaussian field and
\be
\Delta P(k_0,\xv) = 4f_{\rm NL}^{0}\int \frac{d^3 \kv_{\ell}}{\tpc} \phi(\kv_{\ell}) e^{i\kv_{\ell} \cdot \xv}
\ee
is the super cosmic variance contribution. As before, $\Delta P$ may be expanded in multipole moments. 
A spatial modulation of the spectral index was used in \cite{Dai:2013kfa} as an alternative way to produce the power asymmetry. Here we see that such a modulation is a natural consequence of non-Gaussian scenarios that generate a scale-dependent power asymmetry. 

\subsection{Statistical analysis}
The Planck 2013 analysis reported a power deficit at $\ell \lesssim 40$ with a statistical significance of approximately $2.5\sigma-3\sigma$ \cite{Ade:2013kta}, while the more recent Planck 2015 analysis \cite{Aghanim:2015xee} shows a slightly lower statistical significance to the deficit. Although that measurement alone does not require a new model, the scale-dependent non-Gaussian scenario generically generates a power modulation. Therefore, the data should be included in constraining the model. If we include a measurement of the monopole modulation $A_0$ in addition to the dipole power asymmetry amplitude $A$, the combined likelihood for $\fNL$ is:
\be
 \mathcal{L}(\fNL|A, A_0) = p_\chi(A, \sigma) p_{\rm fold}(A_0, \smono).
\ee

\begin{figure}
 \includegraphics[width=0.48\textwidth]{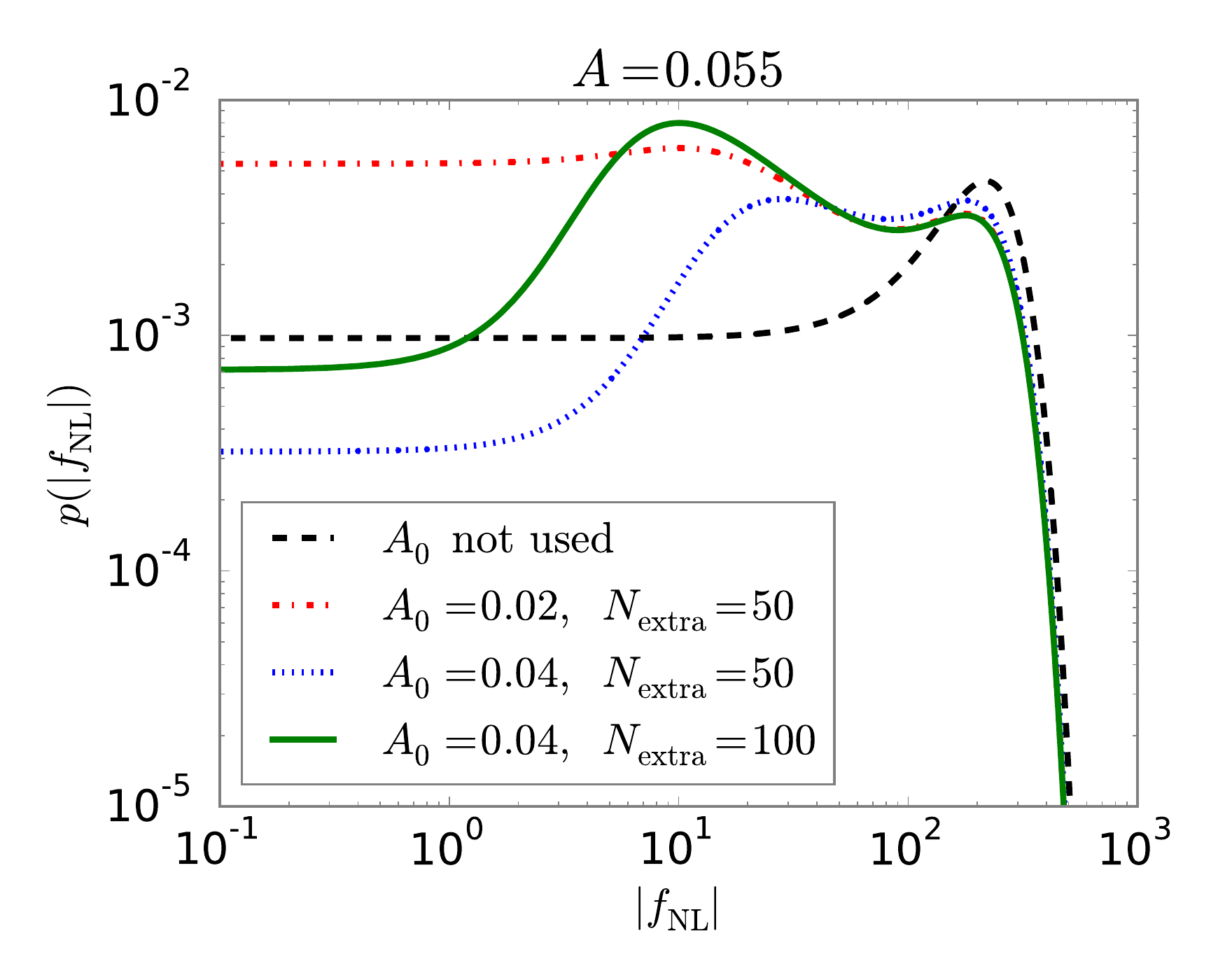}
 \caption{(color online). The posterior distribution of $|\fNL|$ for $A_{\rm obs}=0.055$ and different assumed values of $A_{0, \rm obs}$ and $\Nextra$. Note that the use of $A_0$ changes the shape of the posterior distribution, and generates a peak at around $\fNL\approx10$ for $A_0=0.04, \Nextra=50$ (blue dotted line) in addition to the peak at $|\fNL| \approx 200$ from the use of $A=0.055$ and the bispectrum constraints (black dashed line). For a larger $\Nextra=100$ (green solid line), the peak shifts towards $|\fNL|=0$ as the variance of the prediction of the non-Gaussian model increases; the evidence for the non-Gaussian model also decreases.}
 \label{fig:posteriorA01}
\end{figure}

In Figure \ref{fig:posteriorA01}, we show examples in which, in addition to the power-asymmetry amplitude $A$, we also consider the monopole-modulation amplitude $A_0$. For simplicity, we assume that the result of a scale-dependent non-Gaussianity produces some large (statistically significant at $\approx 3.3 \sigma$, for example) power asymmetry at large scales $\ell\leq100$ and which becomes smaller in magnitude and therefore less significant at small scales.  In Figure \ref{fig:posteriorA01}, therefore, we always take $A=0.055$ and we have also added the estimate of $\fNL \approx -100 \pm100$ at these scales. This simplification allows us to use the Gaussian cosmic variance from our constant $\fNL$ CMB realizations and simply augment the analysis in the previous section with the added information from a possible monopole modulation. This is because we expect the scale-dependent non-Gaussianity to produce a monopole modulation $A_0$ at large scales, the variance of which depends on the strength of non-Gaussianity at the scales used to obtain the monopole modulation amplitude $A_0$. We will define the measured $A_0$ as in Eq.(\ref{eq:A0}) i.e. a weighted average over multipoles $2<\ell \leq 100$. Depending on the values of $\fNL^0$ and $n_{\fNL}$, a scale-dependent non-Gaussianity may generate significant monopole modulation at higher multipoles. But, we will restrict our analysis to a single value $A_0$ obtained from the range $2<\ell \leq 100$. Thus, our simple analysis is only sensitive to the average amplitude of  non-Gaussianity at these scales and cannot constrain $n_{\fNL}$ for which we will need to consider multiple bins of $A_0$ in the data.

For $A_0$ [in Eq.(\ref{eq:A0})] generated at large scales from a scale-dependent non-Gaussianity, $\Cl^{\rm true}$ is the angular power spectrum at large multipoles where the non-Gaussianity and therefore the effect of superhorizon modes on the observed power spectrum is small. In the case of the Planck satellite measurement of $A_0$ in \cite{Ade:2013kta}, the best-fit angular power spectrum, which is also dominated by the larger-$\ell$ modes, is taken as the $\Cl^{\rm true}$.

In Figure \ref{fig:posteriorA01}, we see that for $A_0=0.02$, which is about $1.5\sigma$ in the distribution of $A_0$ values for our Gaussian CMB realizations, the posterior distribution shows more support for the Gaussian model. However, for a larger value $A_0=0.04$ (about $2.9\sigma$), the addition of the $A_0$ data favors the non-Gaussian model for a superhorizon efolds of $\Nextra=50$. For a larger number of superhorizon efolds $\Nextra=100$, the support for the non-Gaussian model again decreases because of the increased variance in the prediction from the non-Gaussian model itself. (However, some care should be taken in extrapolating this result since our analytic expressions are not valid when the non-Gaussianity in the large volume is strong.) The SDDR values for these cases are shown in Table \ref{table:sddrA0}.

\begin{table}[htbp]
  \begin{center}
   \begin{tabular}{|r|c|c|c|}
    
   \hline  $A_{0,\rm obs}$ & $\Nextra$ & SDDR ($B_{01}$) & $\ln{B_{01}}$
\\     \hline \hline
      -  & -  & 0.2012 & -1.603 \\ \hline
      0.02  & 50  & 1.107 & 0.102 \\ \hline
      0.04  & 50  & 0.0664 & -2.712 \\ \hline
      0.04  & 100  & 0.1475 & -1.914 \\ \hline
     \hline
  \end{tabular}
   \end{center}
\caption{SDDR for $A=0.055$ and different observed values of monopole power modulations $A_0$ at large scales. See the text for a discussion on why the evidence increases and decreases for different values of $A_0$ and $\Nextra$.} \label{table:sddrA0}
 \end{table}

We have only shown examples of how using $A_0$ (i.e. power modulation from scale-dependent non-Gaussianity) can change the posterior probability (and hence the evidence) of scale-dependent non-Gaussianity in a simplified situation. A more involved analysis (i.e. including larger $\ell$ modes and measuring the scale dependence of $\fNL$ from bispectrum measurements with the latest data) is necessary to see how useful the inclusion of $A_0$, in addition to the power asymmetry amplitude, proves to be in the actual CMB data. In addition, we have chosen a simple model that may capture the features seen in the data but that may not be easily achievable from an inflationary point of view \cite{Byrnes:2015dub}.

\section{Summary and Conclusion}
\label{sec:conc}
We have performed a systematic study of the power asymmetry expected in the CMB if the primordial perturbations are non-Gaussian and also exist, with no special features, on scales larger than we can observe. We have derived an expression to compute the expected deviations from isotropy of the observed two-point function due to mode coupling with our specific realization of superhorizon modes, which generate anisotropy across our Hubble volume. Although we have focused our analysis on local non-Gaussianity, our method is quite general for describing deviations from statistical isotropy in a finite sub-volume of an isotropic (but non-Gaussian) large volume.

Exploiting the fact that local non-Gaussianity naturally produces a power asymmetry, we have shown how the observed asymmetry can be used for parameter estimation of the non-Gaussian amplitude $\fNL$. We have also combined $\fNL$ constraints from bispectrum and power-asymmetry measurements to evaluate the Bayesian evidence for $\fNL\neq 0$. In our simple examples, we find that the observed CMB power asymmetry only provides weak evidence for non-Gaussianity on large scales.

Many previous works that propose mechanisms to generate the CMB power asymmetry have required a departure from the simplest extrapolation of our observed statistics to larger scales in addition to non-Gaussianity.  The most popular new feature is a large-amplitude superhorizon fluctuation that has an origin beyond inflation.  Generating the observed scale dependence of the power asymmetry requires an additional elaboration, either in the form of scale-dependent non-Gaussianity or isocurvature fluctuations.   However, our results show that scale-dependent non-Gaussianity of the local type is sufficient to generate a dipole power asymmetry at large scales without invoking exotic superhorizon fluctuations.  A value of $\fNL$ consistent with the rather weak large-scale-only constraint (but well above the scale-independent bound) is enough to make the observed power asymmetry no longer anomalous.   It is also worth noting that scale-dependent bispectra are only required for the local ansatz: scale-independent non-local bispectra can generate a scale-dependent power asymmetry.  All that is required to qualitatively match the observed scale dependence of the asymmetry is that smaller scales couple more weakly than large scales do to near-Hubble scale modes.

For scale-dependent non-Gaussianity, we have demonstrated that a large-scale suppression or enhancement of the isotropic power spectrum is expected. Consequently, observations of a scale-dependent shift in the amplitude of the power spectrum can provide additional evidence in favor of non-Gaussianity.   Although we have not found that the observed power deficit at large scales is sufficient to say the data strongly prefer the non-Gaussian model, we have also not performed a thorough analysis nor forecast the improvement possible with future data.  Our parameter estimation and model comparison works were performed in a simplified setting to illustrate that a more careful look at the CMB data is necessary and useful. 

Our results strongly suggest that to evaluate the observational evidence for models of the cosmological perturbations, departures from Gaussianity and from statistical isotropy in the data can and should be considered together. Furthermore, in weighing whether the large-scale CMB anomalies are evidence of something beyond slow-roll inflation lurking just outside the horizon, we should be sure to account for the possibility of non-Gaussian cosmic variance contributions to anisotropy, which do not require the introduction of additional physics.

\acknowledgments
This work is supported by the National Aeronautics and Space Administration under Grant No. NNX12AC99G issued through the Astrophysics Theory Program. Research at Perimeter Institute is supported by the Government of Canada through Industry Canada and by the Province of Ontario through the Ministry of Economic Development \& Innovation. The numerical computations for this work were conducted with Advanced CyberInfrastructure computational resources provided by The Institute for CyberScience at The Pennsylvania State University. 

\appendix
\section{Conventions and Definitions}
\label{app:conventions}
Throughout the paper, we follow the following Fourier convention:
\begin{eqnarray}
 \phi(\xv) &=& \int \frac{d^3\kv}{\tpc} e^{i \kv \cdot \xv} \phi(\kv); \nonumber \\
 \phi(\kv) &=& \int d^3 \xv \; e^{-i \kv \cdot \xv} \phi(\xv).
\end{eqnarray}

As usual, the temperature fluctuations of the CMB are decomposed into spherical harmonics:
\be
 \DeltaTT (\hat{n}) = \sum_{\ell m} a_{\ell m} Y_{\ell m}(\hat{n}),
\ee
with the spherical harmonics normalization given by
\be
 \int d\Omega_{\hat{n}} Y_{\ell m}^*(\hat{n}) Y_{\ell'm'}(\hat{n}) = \delta_{\ell \ell'}\delta_{mm'}.
\ee
Therefore, the multipole coefficients $a_{\ell m}$ of the CMB temperature fluctuations are given by 
\be
a_{\ell m} = \int d\Omega_{\hat{n}} Y_{\ell m}^*(\hat{n}) \DeltaTT (\hat{n}).
\ee

For Sachs-Wolfe temperature fluctuations ($\DeltaTT = - \Phi/3$) and when the Bardeen potential $\Phi$ is a Gaussian field ($\Phi(\xv) = \phi(\xv)$), the Sachs-Wolfe $a_{\ell m}$ is given by
\be
 a_{\ell m} = - \frac{4\pi}{3} i^\ell \int \frac{d^3\kv}{\tpc}  \phi(\kv) \jl(kx) Y_{\ell m}^*(\hat{k}).
\ee

From this, we can obtain the Sachs-Wolfe angular power spectrum $C_\ell$ defined as: $\langle a_{\ell m} a_{\ell'm'}^* \rangle =  \delta_{\ell \ell'}\delta_{mm'} C_\ell$.
\be
 \Cl^{\rm SW} = \frac{4\pi}{9} \int_0^\infty \frac{dk}{k} \jl^2(kx) \mathcal{P}_\phi(k),
\ee
where we have defined the power spectrum $P_\phi(k)$ as
\be
 \langle \phi(\kv) \phi(\kv') \rangle = \tpc \diracdelta(\kv+\kv') P_\phi(k),
\ee
or equivalently, $\langle\phi(\kv) \phi^*(\kv') \rangle = \tpc \diracdelta(\kv -\kv') P_{\phi}(k)$, and the dimensionless power spectrum $\mathcal{P}_\phi$ is defined as $\mathcal{P}_\phi(k) = 2\pi^2 P_\phi(k)/k^3$. 

\section{Derivation of modulated power spectrum for local non-Gaussianity}
\label{app:derivation}
For a homogeneous and isotropic cosmology, the two-point correlation function depends only on the magnitude of the separation between the two points and is the Fourier transform of the power spectrum:
\be
\left \langle \Phi\left(\xv-\frac{\rv}{2}\right) \Phi\left(\xv + \frac{\rv}{2}\right)\right \rangle = \int \frac{d^3 k}{\tpc} P_{\Phi} (k) e^{i\kv\cdot\rv}.
\ee
A hemispherical power asymmetry (a dipole modulation) cannot be described by allowing the power spectrum to be anisotropic in the usual way, $P(k)\rightarrow P(\vec{k})$. (Since the fluctuations are real, the Fourier modes of $\vec{k}$ and $-\vec{k}$ are related, which forbids a dipole modulation. See, e.g., \cite{Pullen:2007tu}). Instead, if the amplitude of the coincident two-point function $\langle\phi(\xv)\phi(\xv)\rangle$ varies spatially, we can likewise introduce position dependence into the power spectrum, $P_{\Phi}(k)\rightarrow P_{\Phi} (k, \xv)$. Both even and odd multipole modulations can now be incorporated.

This inhomogeneous power spectrum is well defined for Fourier modes with wavelengths that are much smaller than the length scale on which the power spectrum changes because these modes have unambiguous and constant wavelengths even if their amplitude varies spatially.  In contrast, a plane wave with an amplitude that varies significantly within one wavelength doesn't have a well-defined amplitude or wavelength.  Therefore, $P_{\Phi} (k, \xv)$ is only defined for wavenumbers such that $P_{\Phi} (k, \xv) \simeq P_{\Phi} (k, \xv+(2\pi/k^2)\kv)$.  On a similar note, defining 
\be
\left \langle \Phi\left(\xv-\frac{\rv}{2}\right) \Phi\left(\xv + \frac{\rv}{2}\right)\right \rangle \equiv \int \frac{d^3 k}{\tpc} P_{\Phi} (k,\xv) e^{i\kv\cdot\rv}
\label{eq:twopoint}
\ee
only makes sense if $P_{\Phi} (k,\xv) \simeq P_{\Phi} (k,\xv\pm\rv)$; otherwise, it is futile to describe the two-point function $\left \langle \Phi\left(\xv-{\rv}/{2}\right) \Phi\left(\xv + {\rv}/{2}\right)\right\rangle$ in terms of a single power spectrum.  

In this appendix we show that a spatially varying power spectrum defined as in Eq.(\ref{eq:twopoint}) arises naturally in non-Gaussian scenarios where short-wavelength modes are coupled to long wavelength modes.
For example, suppose the statistics in some large volume $V_L\rightarrow \infty$ are described by the local ansatz in real space: $\Phi(\xv)= \phi(\xv) + \fNL\left(\phi(\xv)^2 -\langle \phi(\xv)^2\rangle\right)$ where $\phi$ is a Gaussian random field. Fourier modes of the non-Gaussian field are related to those of the Gaussian field by 
\begin{widetext}
\be
\label{app_eq:fnl}
 \Phi(\kv) = \phi(\kv) + \fNL \int \frac{d^3q}{\tpc} [\phi(\kv-\qv) \phi(\qv)-\langle\phi(\kv-\qv) \phi(\qv)\rangle].
\ee
The two-point correlation of the non-Gaussian field is
\begin{eqnarray}
\label{eq:spaceDependent}
\left \langle \Phi\left(\xv-\frac{\rv}{2}\right) \Phi\left(\xv + \frac{\rv}{2}\right)\right \rangle &=& \int \frac{d^3 \kv}{\tpc}  \int \frac{d^3 \kv'}{\tpc} \langle\Phi(\kv) \Phi(\kv')\rangle e^{i(\kv+\kv') \cdot \xv}  e^{i(\kv -\kv')\cdot \frac{\rv}{2}},\\\nonumber
&=&\int \frac{d^3 \kv}{\tpc}  P_\phi(k) e^{i\kv\cdot\rv} \\\nonumber 
&&+\,\fNL \int \frac{d^3 \kv}{\tpc}  \int \frac{d^3 \kv'}{\tpc}e^{i(\kv+\kv') \cdot \xv}  e^{i(\kv -\kv')\cdot \frac{\rv}{2}}\\\nonumber
&&\times\left[\int \frac{d^3 \pv}{\tpc}\langle\phi(\kv')\phi(\pv)\phi(\kv-\pv)\rangle+\int \frac{d^3 \pv'}{\tpc}\langle\phi(\kv)\phi(\pv')\phi(\kv'-\pv')\rangle\right]\\\nonumber
&&+\mathcal{O}(\fNL^2\mathcal{P}_\phi^2)\;.
\end{eqnarray}
\end{widetext}
When we consider statistics entirely within an infinite volume, the terms proportional to a single power of $\fNL$ vanish because $\phi$ is a Gaussian field. In that case, the power spectrum is corrected only by the last term (proportional to $\fNL^2$), which is small when non-Gaussianity is weak. 

Now, suppose we instead consider the statistics in a single sub-volume. Each sub-volume sits on top of a single realization of modes with wavelengths the size of the sub-volume or larger. So, for example, if mode $\pv$ corresponds to a long wavelength mode while modes $\kv$ and $\kv'$ are well within the sub-volume, then 
 \be
 \langle\phi(\kv')\phi(\kv)\phi(\pv)\rangle|_{\rm sub-volume}=\phi(\pv)\langle\phi(\kv')\phi(\kv)\rangle|_{\rm sub-volume}.
 \ee
That is, $\phi(\pv)$ takes a particular value in the sub-volume, while $\phi(\kv)$ and $\phi(\kv')$ are still randomly distributed. Then, taking Eq.(\ref{eq:spaceDependent}) and considering a Fourier mode $\phi(\pv)$ to be stochastic only if $p=|\pv| > k_{\rm min}$:
\begin{widetext}
\begin{eqnarray}
\left \langle \Phi\left(\xv-\frac{\rv}{2}\right) \Phi\left(\xv + \frac{\rv}{2}\right)\right \rangle_{\rm sub-volume} &=& \int_{|\kv|>k_{\rm min}} \frac{d^3 \kv}{\tpc} e^{i\kv\cdot\rv} P_\phi(k) \\\nonumber
&&+\,2\fNL\int \frac{d^3 \kv}{\tpc}  P_\phi(k) e^{i\kv\cdot\rv}\int_{|\pv|<k_{\rm min}} \frac{d^3 \pv}{\tpc}\phi(\pv)[e^{i\pv\cdot(\xv+\rv/2})+e^{i\pv\cdot(\xv-\rv/2)}]\\\nonumber
&&+\,\mathcal{O}(\fNL^2\mathcal{P}_\phi^2),\\\nonumber
&=&\int_{|\kv|>k_{\rm min}} \frac{d^3 \kv}{\tpc} e^{i\kv\cdot\rv} P_\phi(k)\left[1+4\fNL\int_{|\pv|<k_{\rm min}} \frac{d^3 \pv}{\tpc}\left[\phi(\pv)\cos\left(\frac{\pv\cdot\rv}{2}\right)\right]e^{i\pv\cdot\xv}\right].\\
\label{eq:fullPkx}
\end{eqnarray}
\end{widetext}
As discussed above, $P_{\Phi}(k, \xv)$ can be defined in terms of the two-point function $\left \langle \Phi\left(\xv-{\rv}/{2}\right) \Phi\left(\xv + {\rv}/{2}\right)\right\rangle$ only if \mbox{$P_{\Phi}(k, \xv) \simeq P_{\Phi}(k, \xv\pm\rv)$}.  In Eq.(\ref{eq:fullPkx}), we see that the spatial variation of the power spectrum arises from the $e^{i\pv\cdot\xv}$ factor, which implies that $P_{\Phi}(k, \xv)$ will be nearly constant on scales that are shorter than $1/|\pv|$, i.e. scales that are well within the sub-volume.  Therefore, we should restrict the two-point function to separations such that $\pv\cdot\rv\ll1$, in which case the cosine term in Eq.(\ref{eq:fullPkx}) is approximately unity.  Then Eq.(\ref{eq:twopoint}) implies that
\be
 P_{\Phi,S}(k, \xv)  \simeq P_\phi(k)\left[ 1 + 4 \fNL \int \frac{d^3\kv_{\ell}}{\tpc} e^{i \kv_{\ell} \cdot \xv} \phi(\kv_{\ell})\right], \nn \\
\ee
where the subscript $\ell$ specifies that the integral is only over long wavelength modes ($k<k_{\rm min}$).

Expanding the factor $e^{i\kv_l\cdot\xv}$, we can see the effects of non-zero long wavelength modes as a multipole expansion: 
\ba
P_{\Phi,S}(k)&=&P_{\phi}(k)[1+\fNL g_{00}\\\nonumber
&&+\fNL\sum_{M=\{-1,0.1\}}g_{1M}Y_{1M}(\hat{n})+\dots]\;.
\ea
The monopole shift is not observable, so we absorb it into the coefficient to define the observed isotropic power spectrum:
\ba
P\o_{\Phi}(k)&=&P\o_{\phi}(k)\left[1\right.\\
&&\left.+\left(\frac{\fNL}{1+\fNL g_{00}}\right)\sum_{M=\{-1,0.1\}}g_{1M}Y_{1M}(\hat{n})+\dots\right]\;.\nonumber
\ea
Finally, the shift to the power spectrum also shifts the observed value of $\fNL$ as defined from the local template for the bispectrum:
\begin{widetext}
\ba
\langle\Phi_S(k_1)\Phi_S(k_2)\Phi_S(k_3)\rangle&\equiv&(2\pi)^2\diracdelta(\kv_1+\kv_2+\kv_3)B\o(k_1,k_2,k_3);\\\nonumber
B^{\rm local, obs}(k_1,k_2,k_3)&=&2\fNL[1+\fNL g_{00}] P_{\phi}(k_1) P_{\phi}(k_2) + {\rm sym},\\\nonumber
&\equiv&2\fNL\o P\o_{\Phi}(k_1)P\o_{\Phi}(k_2) + {\rm sym}.
\ea
\end{widetext}
Since $P\o_{\Phi}=P_{\phi}[1+\fNL g_{00}]$ (considering only the isotropic piece), the second and third lines imply that $\fNL\o=\fNL/[1+\fNL g_{00}]$. Then, the expression for the power spectrum, including the dipole asymmetry, is
\be
P\o_{\Phi}(k)=P\o_{\phi}(k)\left[1+\fNL\o\sum_{M=\{-1,0.1\}}g_{1M}Y_{1M}(\hat{n})+\dots\right]\;.
\label{eq:dipolePobs}
\ee

The superscript ``obs" on $\fNL$ should also be taken to indicate that the value does {\it not} contain the contribution corresponding to the Maldacena consistency relation for single clock inflation [$\fNL \propto (n_s-1)$], which is unobservable \cite{Mirbabayi:2014hda}.

Related useful works that derive estimators when the primordial temperature field is modulated by an anisotropic field include \cite{Dvorkin:2007jp, Hanson:2009gu}.

\section{Bipolar Spherical Harmonics}
\label{app:biposh}
Many works quantifying the likelihood of the power asymmetry make use of bipolar spherical harmonics \cite{Hajian:2005jh, Hajian:2003qq}. As an aid to that analysis, we repeat the calculation of Appendix \ref{app:derivation}, but for the $a_{\ell m}$. In the Sachs-Wolfe approximation, the statistics of the observed CMB two-point function are simply related to the two-point correlation of the potential at the time of decoupling:
\begin{widetext}
\be
\langle a_{\ell_1m_1}a^*_{\ell_2m_2}\rangle=\frac{1}{9}\int d\Omega_1 Y^*_{\ell_1 m_1}(\hat{n}_1)\int\frac{d^3 \kv_1}{\tpc}e^{i\kv_1\cdot\hat{n}_1x}\int d\Omega_2 Y_{\ell_2 m_2}(\hat{n}_2)\int\frac{d^3 \kv_2}{\tpc}e^{-i\kv_2\cdot\hat{n}_2x}\langle\Phi(\kv_1)\Phi^*(\kv_2)\rangle,
\ee
where $x$ is the comoving distance to the last scattering surface. When the potential is non-Gaussian according to the local ansatz, Eq.(\ref{app_eq:fnl}), the statistics observed in a sub-volume will depend on the realization of the long wavelength modes:
\begin{eqnarray}
\langle a_{\ell_1m_1}a^*_{\ell_2m_2}\rangle_{\rm sub-volume} &=&\frac{1}{9}\int d\Omega_1 d\Omega_2 Y^*_{\ell_1 m_1}(\hat{n}_1)Y_{\ell_2 m_2}(\hat{n}_2)\left[\int_{|\kv|>k_{\rm min}}\int\frac{d^3 \kv_1}{\tpc}P(k_1)e^{i\kv_1\cdot\hat{n}_1x} e^{-i\kv_2\cdot\hat{n}_2x}\right.\\\nonumber
&&+2\fNL\int_{|\kv_1|>k_{\rm min}}\frac{d^3 \kv_1}{\tpc}P(k_1)\int_{|\pv|<k_{\rm min}}\frac{d^3 \pv}{\tpc}\phi^*(\pv)e^{i\kv_1\cdot\hat{n}_1x}e^{-i(\kv_1+\pv)\cdot\hat{n}_2x}\\\nonumber
&&\left.+2\fNL\int_{|\kv_2|>k_{\rm min}}\frac{d^3 \kv_2}{\tpc}P(k_2)\int_{|\pv|<k_{\rm min}}\frac{d^3 \pv}{\tpc}\phi(\pv)e^{i(\pv+\kv_2)\cdot\hat{n}_1x}e^{-i\kv_2\cdot\hat{n}_2x}\right],\\\nonumber
&=&C_{\ell}\delta_{\ell_1\ell_2}\delta_{m_1m_2}+2\fNL(4\pi)(-1)^{m_1}\sum_{L,M}(-i)^L\sqrt{\frac{(2\ell_1+1)(2\ell_2+1)}{4\pi(2L+1)}}C_{-m_1m_2M}^{\ell_1\ell_2L}C_{000}^{\ell_1\ell_2L}\\\nonumber
&&\times\left[C_{\ell_1}\int_{|\pv|<k_{\rm min}}\frac{d^3 \pv}{\tpc}\phi^*(\pv)Y_{LM}(\hat{p})+C_{\ell_2}\int_{|\pv|<k_{\rm min}}\frac{d^3 \pv}{\tpc}\phi(\pv)Y_{LM}(\hat{p})\right].
\end{eqnarray}
Here $C_{\ell}$ is given by Eq.(\ref{eq:Cl}) as usual, and $C_{m_1 m_2 M}^{\ell_1 \ell_2 L}$ are the Clebsch-Gordon coefficients.

The standard notation for the bipolar spherical harmonic expansion is
\be
\langle a_{\ell_1m_1}a^*_{\ell_2m_2}\rangle=C_{\ell}\delta_{\ell_1\ell_2}\delta_{m_1m_2}+\sum_{LM}(-1)^{m_1}C_{-m_1m_2M}^{\ell_1\ell_2L}A_{\ell_1\ell_2}^{LM},
\ee
so that the local model gives
\begin{eqnarray}
A_{\ell_1\ell_2}^{LM}&=&2\fNL(4\pi)(-i)^L\sqrt{\frac{(2\ell_1+1)(2\ell_2+1)}{4\pi(2L+1)}}C_{000}^{\ell_1\ell_2L}\\\nonumber
&&\times\left[C_{\ell_1}\int_{|\pv|<k_{\rm min}}\frac{d^3 \pv}{\tpc}\phi^*(\pv)Y_{LM}(\hat{p})+C_{\ell_2}\int_{|\pv|<k_{\rm min}}\frac{d^3 \pv}{\tpc}\phi(\pv)Y_{LM}(\hat{p})\right].
\end{eqnarray}
\end{widetext}

\section{Notes on numerical realizations}
\label{app:numerics}
In Section \ref{sec:num}, we used Gaussian and non-Gaussian Sachs-Wolfe CMB maps to test some of our analytical formulas. The generation of non-Gaussian CMB maps for a more realistic CMB sky has been described in \cite{Liguori:2003mb, Elsner:2009md}. Following \cite{Elsner:2009md}, we can write the temperature multipole moments as an integration over multipole moments of the Bardeen potential as a function of comoving distance $r$:
\begin{eqnarray}
 a_{\ell m} &=& \int dr \; r^2 \Phi_{\ell m}(r) \alpha_{\ell}(r), \\
 \alpha_{\ell}(r) &=& \frac{2}{\pi} \int dk \; k^2 g_{\ell}(k) \jl(kr),
\end{eqnarray}
where $g_{\ell}(k)$ is the transfer function of temperature in momentum space. The process of generating non-Gaussian CMB maps as detailed in \cite{Elsner:2009md}, therefore, requires generating non-Gaussian $\Phi_{\ell m}$ at different comoving distances $r_i$, considering the covariance $\langle \Phi_{\ell_1 m_1}(r_1) \Phi_{\ell_2 m_2}(r_2) \rangle$, and numerically integrating over the comoving distance $r$. For a Sachs-Wolfe universe, $\g_{\ell}(k) = -\jl(k \rcmb)/3$, and \cite{Komatsu:2001rj}:
\begin{eqnarray}
 \alpha_{\ell}(r) &=& -\frac{2}{3\pi}\int_0^\infty dk\; k^2 \jl(k \rcmb) \jl(kr) \nn \\ &=& -\frac{\diracdelta(r-\rcmb)}{3 \rcmb^2}. \nonumber
\end{eqnarray}
Using this, we obtain:
\be
 a_{\ell m} = -\frac{1}{3} \Phi_{\ell m}(\rcmb).
\ee

Therefore, we can generate the local non-Gaussian CMB Sachs-Wolfe temperature anisotropies simply using Eq.(\ref{eq:localCMB}). The maps used in our study were generated using the HEALPIX software \cite{healpix}, with $\ell_{\rm max}=300$ for the input $C_{\ell}$s and $N_{\rm side}=128$ for map-making. With the 10000 Sachs-Wolfe simulated CMB maps for Gaussian and non-Gaussian potentials, we can perform a number of correlation tests in the Gaussian and non-Gaussian maps to understand what generates the power asymmetry. 

\begin{figure}
 \includegraphics[width=0.48\textwidth]{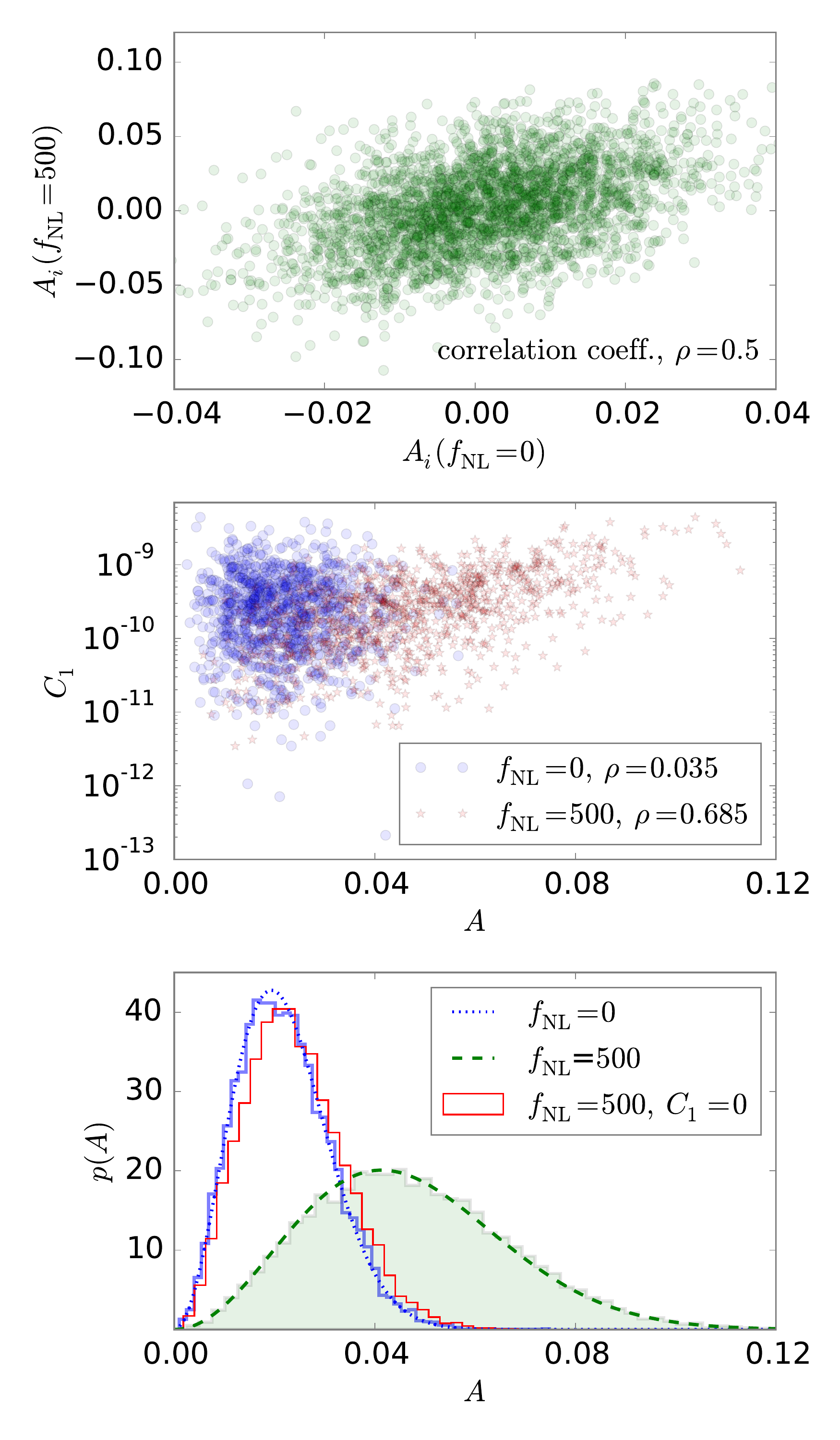}
 \caption{(color online). \textit{Top}: The relation between the $\Cl$ asymmetry in three orthonormal directions for 1000 of our simulated Gaussian maps (x-axis) and the corresponding local non-Gaussian maps with $\fNL=500$ (y-axis). We find a correlation coefficient of 0.5. The correlation coefficient for the corresponding $A$ values (not shown) is 0.25. \textit{Middle}: The correlation between the power asymmetry amplitude and $C_1$ for Gaussian maps (blue circles) and non-Gaussian maps with $\fNL=500$ (red stars). The correlation for the Gaussian case is weak, while the correlation coefficient of the power asymmetry amplitude with $C_1$ in the $\fNL=500$ realizations is strong. This shows that the power asymmetry due to non-Gaussianity also depends on the value of $C_1$, a measure of the background dipole anisotropy for each map. \textit{Bottom}: The power asymmetry amplitude $A$ distribution from Gaussian Sachs-Wolfe CMB maps (blue) and the corresponding power asymmetry amplitude $A$ distribution from non-Gaussian maps (red histogram, the non-Gaussianity is of local type with $\fNL=500$) but with $C_1=0$. For comparison, the green histogram shows the distribution of $A$ with $C_1$ present in the Gaussian maps.}
 \label{fig:corrplots}
\end{figure}

In Figure \ref{fig:corrplots}, we show three plots to illustrate some useful correlations among quantities in the simulated maps. In the top panel of Figure \ref{fig:corrplots}, we plot the directional power asymmetry amplitudes $A_i$ for the Gaussian maps $\fNL=0$ and non-Gaussian maps with $\fNL=500$. The significant correlation between $\fNL=0$ and $\fNL=500$ (with a correlation coefficient, $\rho=0.5$) is expected because both maps contain approximately the same amount of power asymmetry that comes from Gaussian cosmic variance. The $\fNL=500$ CMB skies contain additional power asymmetry on top of the $\fNL=0$ asymmetry. We have checked that the correlation gets weaker for larger values of $\fNL$. 

In the middle panel, we plot $C_1$ (which we have used to model the background dipole anisotropy in density fluctuations) against the dipole asymmetry amplitude $A$ measured in both Gaussian ($\fNL=0$) and local non-Gaussian ($\fNL=500$) models. The $C_1$ values remain almost the same with a correlation coefficient of $0.995$; this indicates that the contribution from the $\mathcal{O}(\fNL^2 \mathcal{P}^2)$ term to the power spectrum is small for $\fNL=500$. We find that the correlation between $C_1$ and $A$ for the Gaussian maps is very weak (correlation coefficient = $0.035$) compared to the correlation in the $\fNL=500$ model (correlation coefficient = $0.685$). This shows that the combination of the background dipole anisotropy and non-Gaussianity is responsible for the power asymmetry in the non-Gaussian maps. To further test this notion, we also generated a set of $\fNL=500$ non-Gaussian maps in which $C_1=0$. As we can see in the bottom panel of Figure \ref{fig:corrplots}, we do not find a significant increase in the power asymmetry distribution $A$ for $\fNL=500$ with $C_1=0$, which provides further evidence that the non-Gaussian power asymmetry is generated by a mode coupling between a background dipole anisotropy and the small-scale modes.

\bibliographystyle{apsrev}
\bibliography{paper} 
\end{document}